\documentclass[preprint,showpacs,preprintnumbers,amsmath,amssymb]{revtex4}

\usepackage{graphicx,psfrag}% Include figure files
\usepackage{dcolumn}% Align table columns on decimal point
\usepackage{bm}% bold math
\newcommand{\nn}{\nonumber}
\def\barr{\begin{array}}
\def\earr{\end{array}}
\def\beq{\begin{equation}}
\def\eeq{\end{equation}}
\def\bea{\begin{eqnarray}}
\def\eea{\end{eqnarray}}
\def\bmath{\begin{displaymath}}
\def\emath{\end{displaymath}}
\def\bq{\begin{quote}}
\def\eq{\end{quote}}

\def\slash#1{\setbox0=\hbox{$#1$}#1\hskip-\wd0\hbox to\wd0{\hss\sl/\/\hss}}

\def\g5{\gamma_5}

\def\Li2{\mbox{Li$_2$}}

\def\lN1{\lambda_{N_1}}

\psfrag{lij}{$\lambda_{ij}$}
\psfrag{th1}{$\theta_1$}
\psfrag{th2}{$\theta_2$}
\psfrag{mN}{$m_N$}
\psfrag{mol}{$|\lambda_{ij}|$}
\psfrag{a}{$a$}
\psfrag{l}{$\lambda_{23}$}
\psfrag{Htm}{$BR(H_0 \to \tau \bar \mu)$}
\psfrag{htm}{$BR(h_0 \to \tau \bar \mu)$}
\psfrag{Atm}{$BR(A_0 \to \tau \bar \mu)$}
\psfrag{meg}{$BR(\mu \to e \gamma)$}
\psfrag{tmg}{$BR(\tau \to \mu \gamma)$}
\psfrag{ttm}{$BR(H_x \to \tau \bar \mu)$}
\psfrag{mth1}{$|\theta_1|$}
\psfrag{mth2}{$|\theta_2|$}
\psfrag{mth3}{$|\theta_3|$}
\psfrag{M0}{$M_0$}
\psfrag{M12}{$M_{1/2}$}
\psfrag{Msu}{$M_{SUSY}$}
\psfrag{lk}{$l_k$}
\psfrag{lm}{$l_m$}
\psfrag{blk}{$\bar {l}_k$}
\psfrag{blm}{$\bar {l}_m$}
\psfrag{s}{$a$}
\psfrag{la}{$\tilde {l}_{\alpha}$}
\psfrag{lb}{$\tilde {l}_{\beta}$}
\psfrag{Hsll}{$BR(H_{SM} \to l_k \bar{l}_m)$}
\psfrag{LBR_SM}{$\log \left(BR(H_{SM} \to l_k \bar{l}_m)\right)$}
\psfrag{LmN}{$\log(m_N)$}
\psfrag{argth1}{$Arg(\theta_1)$}
\psfrag{argth2}{$Arg(\theta_2)$}
\psfrag{argth3}{$Arg(\theta_3)$}
\psfrag{M0M12}{$M_0=M_{1/2}$}
\begin{document}
\preprint{FTUAM-04-16}
\preprint{IFT-UAM/CSIC-04-41}
\preprint{LAPTH-1057/04}
\title{Lepton flavor violating Higgs boson decays\\ from massive seesaw neutrinos}
\author{Ernesto Arganda}
\email{ernesto.arganda@uam.es}
\author{Ana M. Curiel}%
\email{ana.curiel@uam.es}
\author{Mar\'{\i}a J. Herrero}%
\email{maria.herrero@uam.es}
\affiliation{%
Dpto. de F\'{\i}sica Te\'orica, Universidad Aut\'onoma de Madrid, Spain.
}%
\author{David Temes}
\email{temes@lapp.in2p3.fr}
\affiliation{%
Laboratoire de Physique Th\'eorique, LAPTH, France~\footnote{UMR 5108 du CNRS, asoci\'ee \`a l'Universit\'e de Savoie}. 
}%
\begin{abstract}
Lepton flavor violating Higgs boson decays are studied within the context of seesaw models with
Majorana massive neutrinos. Two models are considered: The SM-seesaw, with the Standard
Model Particle content plus three right handed neutrinos, and the MSSM-seesaw, with the Minimal
Supersymmetric Standard Model particle content plus three right handed neutrinos and their
supersymmetric partners. The widths for these decays are derived from
a full one-loop diagrammatic computation in both models, and they are analyzed numerically in
terms of the seesaw parameters, namely, the Dirac and Majorana mass matrices. 
Several possible scenarios for these mass matrices that
are compatible with neutrino data are
considered. In the SM-seesaw case, very small branching ratios are found 
for all studied scenarios. 
These ratios are explained as a consequence of the 
decoupling behaviour of the heavy right handed 
neutrinos. In contrast, in the MSSM-seesaw case, sizable branching
ratios are found for some of the leptonic flavor violating decays of the MSSM neutral Higgs 
bosons and for some choices of the seesaw matrices and MSSM parameters. The
relevance of the two competing sources of lepton flavor changing interactions in the MSSM-seesaw
case is also discussed. The non-decoupling behaviour of the
supersymmetric particles contributing in the loop-diagrams is finally shown.  
\end{abstract}
\maketitle

%%%%%%%%%%%%%%%%%%%%%%%%%%%%%%%%%%%%%%%%%%
\section{\label{sec:Intro} Introduction}
The present strong evidence for lepton flavor changing neutrino 
oscillations~\cite{neutrinodata} 
in solar and atmospheric neutrino data, as well as in the KamLAND reactor 
experiment, implies the existence of non-zero masses 
for the light neutrinos, and provides the first experimental clue for physics beyond the 
Standard Model (SM). The experimentally suggested smallness of the neutrino masses can be
explained in a very simple and elegant way by the seesaw mechanism of neutrino mass generation~\cite{seesaw}. 
This mechanism requires the introduction of heavy right-handed (RH) Majorana neutrinos which
are singlet under the SM gauge symmetry group and whose Majorana masses, $m_{M_i}$, can
therefore be much higher than the SM particle masses. In this context, the smallness of the
light left-handed (LH) neutrino masses appears naturally due to the induced large suppression by
the ratio of the two very distant mass scales that are involved in the seesaw mass matrices,
the Majorana matrix $m_M$ and the Dirac matrix $m_D$. The latter is generated after electroweak symmetry
breaking by $m_D=Y_{\nu}<H>$, where $Y_{\nu}$ is the Yukawa matrix for couplings between the
 RH and LH neutrinos, and  $<H>=v=174$ GeV is the SM
Higgs boson vacuum expectation value. For the one generation case, and assuming
a Yukawa coupling of order one, the suggested small neutrino mass value signals towards a new
phyics mass scale of the order of $m_M \sim 10^{14} \, $ GeV, but the pattern and size of the 
seesaw mass parameters can vary much respect to this in the most general case of three
generations. 

Another appealing feature of the seesaw models is that the RH Majorana neutrinos can 
succesfully generate, through their CP-violating decays and via leptogenesis, the observed 
baryon asymmetry of the Universe. There is, however, one negative aspect in the standard
version of the seesaw models. It is that the presence of the two distant mass scales can lead to a
severe hierarchy problem, and this requires the introduction of supersymmetry (SUSY) 
to be solved. In the SUSY-seesaw models, the hierarchy between $m_M$ and the electroweak scale 
is stabilized  by the new contributions of the SUSY partners of the RH and LH
neutrinos. Thus,
the SUSY-seesaw models, and particularly the simplest version given by the Minimal Supersymmetric
Standard Model (MSSM), are becoming more popular. 

One of the most interesting  features of the SUSY-seesaw models is the associated rich
phenomenology due to the occurrence of lepton flavor violating (LFV) processes. Whereas in the 
standard (non-SUSY) seesaw models the ratios of LFV processes are small due to the smallness of
the light neutrino masses, in the SUSY-seesaw models these can be large due to an important
additional source of lepton flavor mixing in the soft-SUSY-breaking terms. Even in the
scenarios with universal soft-SUSY-breaking parameters at the large
energy scale associated to the SUSY breaking $M_X$ (which could be the Planck mass, the SUSY-GUT mass
or something else, but always well above $m_M$), the running from this scale down to $m_M$ 
induces, via the neutrino Yukawa couplings $Y_{\nu}$, large lepton flavor mixing in the slepton soft masses, 
and provides the so-called slepton-lepton misalignment, which in turn
generates 
non-diagonal lepton flavor interactions. 
%For instance, in the MSSM-seesaw
%version, there 
%appear interactions of the type $\rm{chargino-(lepton)_i-(sneutrino)_j}$ and 
%$\rm{neutralino-(lepton)_i-(slepton)_j}$, with $\rm{i\neq j}$,
These interactions can induce sizable ratios in several LFV processes with SM charged leptons in the 
external legs, which are actually being tested experimentally with high precision and therefore provide 
a very interesting window to look for indirect SUSY signals.

In addition to the previously mentioned radiatively induced LFV effect,
there is
another source of radiative LFV effects in the MSSM-seesaw, which is generated 
from 
the neutrino sector mixing. Namely, the off-diagonal entries 
of the Maki-Nakagawa-Sakata (MNS) matrix, $U_{MNS}$~\cite{MNS}, produce 
lepton flavor non-diagonal interactions involving neutrinos which could also 
generate, 
via loops, important contributions to the LFV processes 
with SM leptons in the external legs. 
%,
%as for instance, $\rm{chargino-(lepton)_i-(sneutrino)_j}$, 
%$\rm{H^{\pm}-(lepton)_i-(neutrino)_j}$, 
%and $\rm{W^{\pm}-(lepton)_i-(neutrino)_j}$ and others, with $\rm{i\neq j}$w%which  
%can generate, via loops, important contributions to the mentioned LFV proces%ses with SM
%charged leptons in the external legs. 
%
%Notice, that the third interaction appears 
%in both the SM-seesaw and MSSM-seesaw versions,  whereas the first and the second ones 
%appear only 
%in the MSSM-seesaw (the second one would also appear in a 2HDM-seesaw version, but we do not
%consider this here). Given the present experimental evidence for large off-diagonal entries in the MNS 
%matrix, more specifically those regarding the neutrino mixings between the third and second 
%generations and between the first and second ones, this source of LFV effects should not be
%ignored. 
In the MSSM-seesaw, these LFV effects from neutrino mixing could indeed compete in some
processes with the other source
from slepton-lepton misalignment. 
 This competitivity of the two flavor changing sources 
 has indeed been noticed previously in the 
squark sector, concretely in the SUSY electroweak one-loop contributions to flavor changing Higgs boson decays,
$h^0\rightarrow \bar{b}s,\bar{s}b$~\cite{Curiel:2003uk}. Interestingly, 
the lepton sector could manifest more evidently these  
competing effects than the quark sector, since the MNS matrix being prefered by neutrino data 
is clearly non-diagonal, whereas the CKM matrix is close
to the identity matrix.

Among the various LFV processes that have been considered in the literature, the most 
fruitfull ones are the radiative $\mu \rightarrow e \gamma$, $\tau \rightarrow \mu \gamma$,  
and $\tau \rightarrow e \gamma$ 
decays~\cite{raredecays,Casas:2001sr,Chankowski:2004jc,Pascoli:2003rq}, 
since their branching 
ratios are tested with high precission~\cite{taumu,taue,mue}. These usually provide the most
restrictive experimental bounds on the MSSM-seesaw parameters. Another 
interesting LFV decays  include $\tau$ rare decays, $Z$ boson decays 
$Z\rightarrow l_k\bar{l}_m$ and Higgs boson decays $H\rightarrow  l_k\bar{l}_m$,
with $k\ne m$.

We are interested here in the LFV Higgs boson Decays (LFVHD), 
$H\rightarrow \tau\bar{\mu},\tau\bar{e}, \mu\bar{e}$, and 
the branching ratios that can be generated in the context of the seesaw models 
with parameters being compatible with the neutrino data and the most relevant 
data of $\tau $ and $\mu$ radiative decays.  In this work we consider both versions 
of the seesaw mechanism for neutrino mass generation:  the SM-seesaw 
and the MSSM-seesaw. In the first one, the SM particle content is enlarged by 
three singlet RH Majorana neutrinos, and in the second one, the MSSM particle
content is enlarged by three singlet RH Majorana neutrinos and their
corresponding SUSY partners. We present here a complete one-loop
computation of the LFVHD widths in both seesaw scenarios and analyze 
the size of the associated branching ratios in terms of the seesaw parameters.
 In order to make contact between the seesaw mass matrices and the experimental
 neutrino data, we use the general parametrization introduced in ref.~\cite{Casas:2001sr},
 where the Yukawa neutrino couplings, and therefore $m_D$, are expressed in terms of the 
 three physical light
 neutrino masses, $m_{\nu_i}$, the three physical heavy neutrino masses, $m_{N_i}$, 
 the $U_{MNS}$ matrix, and a general 
 complex $3 \times 3$ orthogonal matrix $R$. 
 
 The LFVHD  within the SM-seesaw was studied some time ago in 
 ref.~\cite{Pilaftsis1}. 
 There, it was considered a very specific seesaw scenario where all the light 
 neutrinos were exactly massless at tree level~\cite{Pilaftsis2}
 and the Dirac mass was taken very large. The conclusion was 
 that branching ratios for $H_{SM}\rightarrow \tau \bar{\mu}$ decays as large as  
 $10^{-4}-10^{-5}$ can be achived for $m_{H_{SM}}\leq 140 \, $ GeV. Besides, 
 these ratios were found to grow with the heavy neutrino masses, 
 and this growing suggested a non-decoupling behaviour of the heavy
 neutrinos. The crutial assumption for that behaviour was to take very large 
 $m_D$ values, which implied very strong neutrino Yukawa couplings. Here, we have recalculated the LFVHD rates in the complete one-loop
 diagrammatic approach and updated the numerical estimates in the light of the 
 recent neutrino data, by fixing the input parameters  $m_{\nu_i}$ and 
 $U_{MNS}$ to their data prefered values. 
  We then make our estimates for
 specific choices of the input unknown parameters, $m_{N_i}$ and $R$ and pay special atention to 
 those 
 which generate succesfull  baryon asymmetry. 
 For all the studied cases we find ratios that are many
 orders of magnitude smaller than in ref.~\cite{Pilaftsis1}. The reason 
 for these small
 ratios is because the heavy RH neutrinos do indeed decouple in our SM-seesaw
 scenario and our assumptions for the seesaw parameters do not imply large $m_D$ 
 values. 
For completeness, we also include in this part, an estimate of the LFVHD rates in the case of 
Dirac massive neutrinos and compare them with the previous estimate of ref.~\cite{Diaz-Cruz:1999xe}.

 The second part of this work concerns with the evaluation of the LFVHD ratios in the context of the
 MSSM-seesaw. Concretely, $h^0,H^0,A^0\rightarrow l_k\bar{l}_m$, with $k\neq m$. 
 The subject of LFVHD being generated from loops of SUSY particles has been considered previously 
 in refs.~\cite{Diaz-Cruz:1999xe} and~\cite{Brignole:2003iv}. In
 ref.~\cite{Diaz-Cruz:1999xe} it was analyzed a specific SUSY-SU(5)
 scenario  where the slepton-lepton misalignment was generated exclusively from the running of the 
 trilinear A-terms. 
 On the other hand, the computation of ref.~\cite{Brignole:2003iv} was not in the context of 
 the MSSM-seesaw but in a
 more generic scenario for slepton-lepton misalignement. Besides, in ref.~\cite{Brignole:2003iv}, the effective lagrangian 
 approach that
 is valid for large $\tan\beta$ values and large SUSY mass values is used. We present 
 here instead, a complete one-loop computation in the MSSM-seesaw context and 
 do not rely on any of the above
 approximations. We include in the
 computation both sources of lepton flavor violating interactions, the slepton-lepton misalignment
 and the neutrino mixing via $U_{MNS}$. The slepton-lepton misalignment is generated, as it is
 usual in the seesaw models, by the renormalization group running 
  of the slepton soft parameters from the high energies $M_X$ down to $m_M$. The 
  diagonalization of the  generated
  slepton mass matrix is then performed. 
  That is, we do not rely on either the mass insertion approximation
  or the large $\tan\beta$ effective lagrangian approach and, therefore, our results
 are  valid for all $\tan\beta$ values and all soft-SUSY-breaking mass values. We will
 explore the size of the branching ratios for the Higgs decays as a function of the relevant MSSM
 parameters, which within the context of mSUGRA are the universal soft masses $M_0$, $M_{1/2}$
and $\tan\beta$,  and of the relevant seesaw parameters, which are $m_{N_i}$ and the $R$ matrix.
We will analyze in parallel the branching ratios for 
  the $l_j\to l_i \gamma$ decays as a function of the same parameters. The requirement of compatibility 
  with the present data on 
  $l_j\to l_i \gamma$ decays, mainly $\mu \to e \gamma$ and $\tau \to \mu \gamma$, will provide us with the maximum allowed ratios 
  for the Higgs decays.
 We will also study the behaviour of the LFVHD widths in the limit of very heavy SUSY 
 masses and will find that the sleptons, sneutrinos, charginos and neutralinos do not decouple in
 this observable. For large SUSY masses, large $\tan\beta$ and particular choices of the see-saw
 parameters we will find agreement with the numerical results of ref.~\cite{Brignole:2003iv}.

 The paper is organized as follows.  Section II contains a short summary of the mass parameters and
 mixings in the neutrino sector of the seesaw models. The relation between these parameters 
 and the experimental neutrino masses and mixings is also 
 included. Section III is devoted to the computation and analysis of the LFVHD rates 
 in the context of the SM-seesaw. The decoupling behaviour of the
 heavy neutrinos is also studied in this section. Section IV starts by presenting the two sources of LFV interactions
 in the context of the MSSM-seesaw. Next the computation and analysis 
 of the LFVHD rates within that context is included. The non-decoupling behaviour of the SUSY particles in the LFVHD
 widths is studied at the end of this section. Section V is devoted to the conclusions.     

%%%%%%%%%%%%%%%%%%%%%%%%%%%%%%%%%%%%%%%%%%

%%%%%%%%%%%%%%%%%%%%%%%%%%%%%%%%%%%%%%%%%%%%%%%%%%%%%%%%%%%%%%%%%%%%%%%%%%%
\section{\label{sec:2} Neutrino masses and mixings in the seesaw models}
In this section and in order to fix our notation we briefly review the mass 
parameters and  mixings in the neutrino sector of the seesaw models and relate 
them to the physical light neutrino masses and neutrino mixing angles which are extracted 
from neutrino data. We follow closely here and in the next sections 
the notation of refs.~\cite{Pilaftsis1,Pilaftsis2} for SM-seesaw and ref.~\cite{Casas:2001sr} for SUSY-seesaw and for the 
connection with neutrino data. 

We start with the Yukawa-sector of the SM-seesaw that contains the three 
LH SM neutrinos $\nu_{L, i}^o$ and three extra RH massive neutrinos 
$\nu_{R, i}^o$, whose Yukawa interactions provide, after spontaneous electroweak 
symmetry breaking, together with the right handed neutrino masses, the following mass Lagrangian containing the Dirac and Majorana mass terms,
\begin{equation}
-L^\nu_{mass} = \frac{1}{2} (\overline{\nu^0_L}, (\overline{\nu^0_R})^C)
M^\nu \left(\barr{c} (\nu^{0}_L)^C\\ \nu^0_R \earr \right)\  + h.c.,
\end{equation} 

where,
\begin{equation}
M^\nu\ =\ \left( \barr{cc} 0 & m_D\\ m_D^T & m_M \earr \right).
\label{mass6x6}
\end{equation}  

Here $m_D$ is the $3 \times 3$ Dirac mass matrix that is related to the $3 \times 3 $ Yukawa 
coupling matrix $Y_{\nu}$ and the SM Higgs vacum expectation value, $<H>=v=174$ GeV, by 
$m_D=Y_{\nu} <H>$; and $m_M$ is the $3 \times 3$ Majorana mass matrix for the RH massive neutrinos
 that is real, non singular and symmetric.

The mass matrix $M^{\nu}$ is a $6 \times 6$ complex symmetric matrix that can be diagonalized by a $6 \times 6$ unitary matrix $U^{\nu}$ in the following way:

\begin{equation}
U^{\nu T}M^\nu U^\nu =\hat{M}^\nu = diag (m_{\nu_1},m_{\nu_2},m_{\nu_3},m_{N_1},m_{N_2},m_{N_3}).
\label{matrizU}
\end{equation}
This gives 3 light Majorana neutrino mass eigenstates $\nu_i$, with masses $m_{\nu_i}$ (i=1,2,3), and three heavy ones $N_i$, with masses $m_{N_i}$ (i=1,2,3), which are related to the weak eigenstates via,

\begin{equation}
\left(\barr{c} \nu^0_L \\ (\nu^{0}_R)^C \earr \right)\ =\ 
U^{\nu\ast}\ \left(\barr{c} \nu_L \\ N_L \earr \right)\quad \mbox{and}\quad
\left(\barr{c} (\nu^{0}_L)^C \\ \nu^0_R \earr \right)\ =\ 
 U^\nu\ \left(\barr{c} \nu_R \\ N_R \earr \right).
\end{equation}
The seesaw mechanism for neutrino mass generation assumes a large separation between the two 
mass scales involved in $m_D$ and $m_M$ matrices. More specifically, 
we shall assume here that all matrix elements of $m_D$ are much smaller than those 
of $m_M$, $m_D<<m_M$, and we will perform an analytical expansion of all relevant 
interaction parameters and observables in power series of a matrix defined as,
\begin{eqnarray}
\xi &\equiv &m_D m_M^{-1}.
\end{eqnarray}
 In particular, the previous diagonalization of the mass matrix $M^{\nu}$ can be solved 
 in power series of $\xi$~\cite{Grimus:2000vj}. For simplicity, we choose to work here and in the rest
  of this paper, in a flavor basis where the RH Majorana mass matrix, $m_M$, and the 
  charged lepton mass matrix, $M^l$, are flavor diagonal. This means that all flavor mixing of 
  the light sector is included in the mixing matrix $U_{MNS}$. By working to the lowest 
  orders of these power series expansions one finds, in the flavor basis, the following neutrino $3 \times 3$ matrices, 

\begin{eqnarray}
m_{\nu}&=&-m_D \xi^T + \mathcal{O}(m_D \xi^3) \simeq -m_D m_M^{-1}m_D^T\\ \nn
m_N &=& m_M + \mathcal{O}(m_D \xi) \simeq m_M.
\end{eqnarray}
Here, $m_N$ is already diagonal, but $m_{\nu}$ is not yet diagonal. The rotation from this flavor
 basis to the mass eigenstate basis is finally given by the MNS unitary matrix, $U_{MNS}$. 
 Thus, 
\begin{eqnarray}
m_{\nu}^{diag}&=&U_{MNS}^T m_{\nu} U_{MNS}= diag(m_{\nu_1},m_{\nu_2},m_{\nu_3}),\\ \nn
m_N^{diag} &=& m_N = diag(m_{N_1},m_{N_2},m_{N_3}),
 \end{eqnarray}
and the diagonalization of $M^{\nu}$ in eqs.~(\ref{mass6x6}) and (\ref{matrizU}) can be performed by the following unitary $6 \times 6$ matrix:

\begin{equation}
U^\nu\ =\ \left( \barr{cc} (1-\frac{1}{2} \xi^* \xi^T) U_{MNS} & \xi^* (1-\frac{1}{2} \xi^T \xi^*)\\ -\xi^T (1- \frac{1}{2} \xi^* \xi^T)U_{MNS} & (1-\frac{1}{2} \xi^T \xi^*) \earr \right) + \mathcal{O}(\xi ^4).
\label{eq8}
\end{equation}
As for the $U_{MNS}$ matrix, we use the standard parametrization given by,

\begin{equation}
U_{MNS}\ =\ \left( \barr{ccc} c_{12} c_{13} & s_{12} c_{13}& s_{13} e^{-i \delta}\\ -s_{12} c_{23}-c_{12}s_{23}s_{13}e^{i \delta} & c_{12} c_{23}-s_{12}s_{23}s_{13}e^{i \delta} & s_{23}c_{13} \\ s_{12} s_{23}-c_{12}c_{23}s_{13}e^{i \delta} & -c_{12} s_{23}-s_{12}c_{23}s_{13}e^{i \delta} & c_{23}c_{13}\earr \right) diag(1,e^{i \alpha},e^{i \beta}).
\label{Umns}
\end{equation}
where $c_{ij} \equiv \cos \theta_{ij}$ and $s_{ij} \equiv \sin \theta_{ij}$.

Finally, in order to  make contact with the experimental data, we use the method proposed 
in ref~\cite{Casas:2001sr}. It provides a simple way to reconstruct the Dirac mass matrix by 
starting with the physical light and heavy neutrino masses, the $U_{MNS}$ matrix, and 
a general complex and orthogonal 
matrix R. With our signs and matrix conventions this relation can be written as~,
\begin{equation}
m_D^T =i m_N^{diag \, 1/2} R m_{\nu}^{diag \, 1/2} U_{MNS}^+
\label{Rcasas}
\end{equation}
where, $R^T R=1$.
 Thus, instead of proposing directly possible textures for $m_D$, 
one proposes possible values for $m_{N_1} \, ,m_{N_2} \, ,m_{N_3} $ and R, and 
sets $m_{\nu_1} \, ,m_{\nu_2} \, ,m_{\nu_3} $ and $U_{MNS}$ to their suggested 
values from the experimental data. Notice 
that any hypothesis for R different from the unit matrix
 will lead to an additional lepton flavor mixing between the LH and RH 
 neutrino sectors. Notice also that the previous relation holds at 
 the energy scale $m_M$, and to use it properly one must use the RGE to run 
 the input experimental data $m_{\nu}^{diag}$ and $U_{MNS}$ from the low 
 energies $m_W$ up to $m_M$. We have computed here these running effects by
 solving the RGE in the one loop approximation~\cite{RGE}, and we have 
considered the corresponding  radiatively corrected neutrino masses and 
$U_{MNS}$ matrix elements in our computation of the LFV rates. 

For the numerical estimates in this paper we will consider the following two 
plausible scenarios, at the low energies, being compatible with data:
\begin{itemize}
\item Scenario A:
quasi-degenerate light and degenerate heavy neutrinos,
\begin{eqnarray}
m_{\nu_1}&=&0.2 \, eV \, , m_{\nu_2}=m_{\nu_1}+\frac{\Delta m_{sol}^2}{2 m_{\nu_1}} \, , m_{\nu_3}=m_{\nu_1}+\frac{\Delta m_{atm}^2}{2 m_{\nu_1}}, \\ \nn
m_{N_1} &=& m_{N_2}= m_{N_3}= m_N
\end{eqnarray}
\item Scenario B:
hierarchical light and hierarchical heavy neutrinos,
\begin{eqnarray}
m_{\nu_1} &\simeq& 0 \, eV \, , m_{\nu_2}= \sqrt{\Delta m_{sol}^2} \, , m_{\nu_3}=\sqrt{\Delta m_{atm}^2}, \\ \nn
m_{N_1} &\leq & m_{N_2} < m_{N_3}
\end{eqnarray}
%\item Scenario C:
%hierarchical light and degenarate heavy neutrinos,
%\begin{eqnarray}
%m_{\nu_1} &\simeq& 0 eV \, , m_{\nu_2}= \sqrt{\Delta m_{sol}^2} \, , m_{\nu_3}=\sqrt{\Delta m_{atm}^2}, \\ \nn
%m_{N_1} &\simeq& m_{N_2} \simeq m_{N_3}
%\end{eqnarray}

\end{itemize}

In the two above scenarios, we will fix the input low energy data to the following values, 
$\sqrt{\Delta m_{sol}^2}=0.008$ eV, $\sqrt{\Delta m_{atm}^2}=0.05$ eV, 
$\theta_{12}=\theta_{sol}=30^o$, $\theta_{23}=\theta_{atm}=45^o$, 
$\theta_{13}=0^o$ and $\delta = \alpha= \beta =0$ 
(See for instance, ref.~\cite{review}). 

Regarding the matrix R, we will consider correspondingly one of the following three cases:
\begin{itemize}
\item Case 0
\begin{equation}
R=R_0=1 
\end{equation}
This is a reference case which is chosen here just because it is the simplest possibility.
\item Case 1 
\begin{equation}
R=R_{1}\ =\ \left( \barr{ccc} c_{2} c_{3} 
& -c_{1} s_{3}-s_1 s_2 c_3& s_{1} s_3- c_1 s_2 c_3\\ c_{2} s_{3} & c_{1} c_{3}-s_{1}s_{2}s_{3} & -s_{1}c_{3}-c_1 s_2 s_3 \\ s_{2}  & s_{1} c_{2} & c_{1}c_{2}\earr \right).
\end{equation}
where $c_i\equiv \cos \theta_i$, $s_i\equiv \sin\theta_i$ and
$\theta_1$, $\theta_2$ and $\theta_3$ are arbitrary complex angles.
 This parametrization was proposed in ref.~\cite{Casas:2001sr} for the study of $\mu \rightarrow e
 \gamma$ decays. It has also been considered in ref.~\cite{Chankowski:2004jc} with specific values for 
 the $\theta_i$ angles to study the implications for baryogenesis in the case of
 hierarchical neutrinos.
\item Case 2 
\begin{equation}
R=R_2=e^{i A} O,
\end{equation}
 with $O=1$, and
\begin{equation}
A\ =\ \left( \barr{ccc} 0 & a & b\\ -a & 0 & c \\ -b & -c & 0 \earr \right).
\end{equation}
Here, $a$, $b$, and $c$ are three real parameters that are constrained by perturbativity of
the Yukawa couplings. In particular, for $a=b=c\equiv k$ and $m_{\nu_{1,2,3}}\simeq 0.2eV$, 
it leads to $k < (1.4,0.9,0.3)$ for $m_{N_{1,2,3}}\simeq (10^{10},10^{12},10^{14})$ GeV respectively.   
This choice has been proven in ref.~\cite{Pascoli:2003rq} to provide successful baryon asymmetry via leptogenesis for the case of quasi-degenerate neutrinos. 
\end{itemize}
All that has been sumarized in this section applies to the  MSSM-seesaw model as well. 
The only difference is that in this case, $m_D$ is related to one of the two Higgs vacum 
expectation values by $m_D = Y_{\nu} <H_2>$, where $ <H_2>=v_2= v \sin \beta$.  
%%%%%%%%%%%%%%%%%%%%%%%%%%%%%%%%%%%%%%%%%%%%%%%%%%%%%%%%%%%%%%%%%%%%%%%%%%%%%%%
\section{\label{sec:3} Lepton Flavour Violating Higgs Decays in the SM-Seesaw}
In this section we compute and analyze  the LFVHD widths, 
$\Gamma(H\rightarrow l_k\bar{l}_m)$ with $k \neq m$, within the context of the SM-seesaw 
with three RH neutrinos which has been shortly reviewed in the previous section. 
The branching ratios for these decays were studied some time ago in ref.~\cite{Pilaftsis1} in 
a particular scenario of the SM-seesaw where all the light neutrino masses were 
exactly zero at the tree level and the Dirac mass was taken very large. 
The small masses for the light neutrinos were then 
generated by the one-loop electroweak radiative corrections~\cite{Pilaftsis2}. 
Besides, the numerical results of ref.~\cite{Pilaftsis1} for the case of 3 generations 
were obtained under particular approximations for the evaluation of the one-loop diagrams, 
as for instance, $m_{l_1}=0$, $\frac{m_{l_2}^2}{m_W^2}<<1$, $\frac{m_H^2}{4m_W^2}<<1$, 
the latter being nowadays clearly not a very good approximation given the present lower 
bound on $m_H$ from LEP of about 115 GeV. Here we have recomputed these LFVHD widths for 
all the channels and we have  included all contributing one-loop diagrams 
without assuming any approximation. One of the main points of our analysis is 
the update of the numerical results by taking into account the present experimental neutrino data.
 In addition we will reanalyze the behaviour of these partial widths with the large heavy neutrino 
 masses. In contrast to ref.~\cite{Pilaftsis1}, where the important enhancement of the widths 
 found with $m_{N_i}$  suggested a non-decoupling behaviour of the heavy neutrinos,  
 we will find instead a clear decoupling behaviour. The main difference between 
 ref.~\cite{Pilaftsis1} and us is the assumption on $m_D$,  which in their work was taken 
 very large (therefore, leading to strong neutrino Yukawa couplings), whereas in our case
 the assumptions for the input $m_{N_i}$ and $R$, do not lead to large $m_D$.   

We start by writting down the interactions that are relevant for the computation 
of the LFVHD widths 
to one-loop and in the mass eigenstate basis. Denoting the Majorana neutrino mass eigenstates 
collectively by $n_i$ (i.e., $n_i \equiv \nu_i$ for i=1,2,3 and $n_i=N_{i-3}$ for i=4,5,6), 
the relevant interactions between $n_i$ and $W^{\pm}$, H and the Goldstone bosons $G^{\pm}$ can be written respectively as follows,

\begin{eqnarray}
\mathcal{L}_{int}^{W^{\pm}} &=& 
\frac{-g}{\sqrt{2}} W^{\mu -}\bar{l_i} B_{l_i n_j} \gamma_{\mu} P_L n_j + h.c,\nonumber\\ 
\mathcal{L}_{int}^{H} &=& 
\frac{-g}{4 m_W} H \bar{n_i} \left[ (m_{n_i}+m_{n_j}) Re(C_{n_i n_j})+i \gamma_5 (m_{n_j}-m_{n_i}) Im(C_{n_i n_j})\right] n_j,\nonumber\\ 
\mathcal{L}_{int}^{G^{\pm}} &=& 
\frac{-g}{\sqrt{2} m_W} G^{\mu -}\left[\bar{l_i} B_{l_i n_j} (m_{l_i} P_L - m_{n_j} P_R) n_j \right] + h.c
\end{eqnarray}

where, the coupling factors $B_{l_i n_j}$ (i=1,2,3, j=1,..6) and $C_{n_i n_j}$ (i,j=1,...6) are defined in terms of the $U^{\nu}$ matrix of eq.~\ref{matrizU} by:

\begin{eqnarray}
B_{l_i n_j} = U_{ij}^{\nu *}\\
C_{n_i n_j} = \sum_{k=1}^3 U_{k i}^{\nu} U_{kj}^{\nu *}
\end{eqnarray}

Notice that our particular choice of diagonal charged leptons in the flavor basis is equivalent to 
assume a $V$ matrix in~\cite{Pilaftsis1} equal to the identity matrix. The expansions of these coupling matrices in 
power series of the $\xi$  matrix can be easily derived from the expansion of $U^{\nu}$ 
in eq.~\ref{eq8}. For brevity, we omitt here the indices and use a self-explanatory short notation. They are given by,

\begin{equation}
B_{l n} = (B_{l \nu}, B_{l N}),
\end{equation}

\begin{equation}
C_{n n} \ =\ \left( \barr{cc} C_{\nu \nu} & C_{\nu N}\\ C_{N \nu} & C_{N N}\earr \right).
\end{equation}

where,

\begin{eqnarray}
B_{l \nu}&= &(1-\frac{1}{2} \xi \xi^+) U_{MNS}^*+ \mathcal{O} (\xi^4) \nonumber \\
B_{l N}&=& \xi (1-\frac{1}{2} \xi^+ \xi) + \mathcal{O} (\xi^5) \nonumber \\
C_{\nu \nu}&= &U_{MNS}^T (1- \xi \xi^+)U_{MNS}^*+ \mathcal{O} (\xi^4) \nonumber \\
C_{\nu N}&=& C_{N \nu}^+= U_{MNS}^T \xi (1- \xi^+ \xi) + \mathcal{O} (\xi^5) \nonumber \\
C_{N N}&= &\xi^+ \xi + \mathcal{O} (\xi^4).
\label{couplingmatrices}
\end{eqnarray}

After the computation of the 10 contributing one-loop diagrams, drawn in fig.~\ref{fig:1}, 
we find the analytical results presented in Appendix A which have been written in terms of 
the standard one-loop integrals, $C_0, B_0, C_{12},...$ etc, whose definitions can be found for
instance in~\cite{Hollik}. These provide the total contributions to the relevant form factors $F_L$ 
and $F_R$ that are related to the decay amplitude F by:

\begin{equation}
i F = -i g \bar{u}_{l_k} (-p_2) (F_L P_L + F_R P_R) v_{l_m}(p_3)
\end{equation}

where, 
\begin{equation}
F_L = \sum_{i=1}^{10} F_L^{(i)},\,\,  F_R = \sum_{i=1}^{10} F_R^{(i)}
\end{equation}
 and $p_1=p_3-p_2$ is the ingoing Higgs boson momentum. 
 
 The LFVHD widths can be obtained finally from these form factors by:

\begin{eqnarray}
\Gamma (H \to {l_k} \bar{l}_m)& = &\frac{g^2}{16 \pi m_H} 
\sqrt{\left(1-\left(\frac{m_{l_k}+m_{l_m}}{m_H}\right)^2\right)
\left(1-\left(\frac{m_{l_k}-m_{l_m}}{m_H}\right)^2\right)}. \nonumber\\
&&.\left((m_H^2-m_{l_k}^2-m_{l_m}^2)(|F_L|^2+|F_R|^2)- 4 m_{l_k} m_{l_m} Re(F_L F_R^*)\right)
\label{decay}
\end{eqnarray}

Notice that since we will consider complex R matrices, the corresponding decay widths for the CP conjugate 
states, in general, can be different. We do not study here these CP conjugate 
decays and concentrate on 
the $H\rightarrow \tau\bar{\mu},\tau\bar{e}, \mu\bar{e}$ decays.

Regarding the analytical results, it is worth mentioning that we have checked the finiteness 
of the total form factors $F_L$ and $F_R$. When summing up all involved indices in each diagram, 
we find out, in agreement with ref.~\cite{Pilaftsis1}, that the only divergent diagrams left are 1, 8 and 10 
and these divergences cancel among each other, providing the expected finite result. 

 We next present the numerical results. The one-loop functions 
$C_0, B_0,..$ have been evaluated with the Mathematica package of~\cite{loops}. 
For the numerical estimates of the branching ratios we evaluate 
the coupling matrices of  eq.~\ref{couplingmatrices}
in terms of $\xi$, for the particular input values of $m_D$ and 
$m_M$ that are compatible with the neutrino data. That is,  we get $m_D$ from 
eq.~\ref{Rcasas}, for the 
two scenarios A and B and for various choices of the input masses 
$m_{N_{1,2,3}}$ 
and the matrix R, 
as explained in sec.~\ref{sec:2}, and take $m_{M_i}=m_{N_i}$. The total width
has been evaluated with the HDECAY programme~\cite{Djouadi:1997yw}.
%\vspace{1cm}
%%%%%%%%%%%%%%%%%%%%%%%%%%%%%%%%%%%%%%%%%%%%%%%%%%%%%%%%%%%%%%%%%%%%%%%%%%%%%%%%%%%%%%%

\begin{figure}
\includegraphics[width=15.0cm,height=15cm]{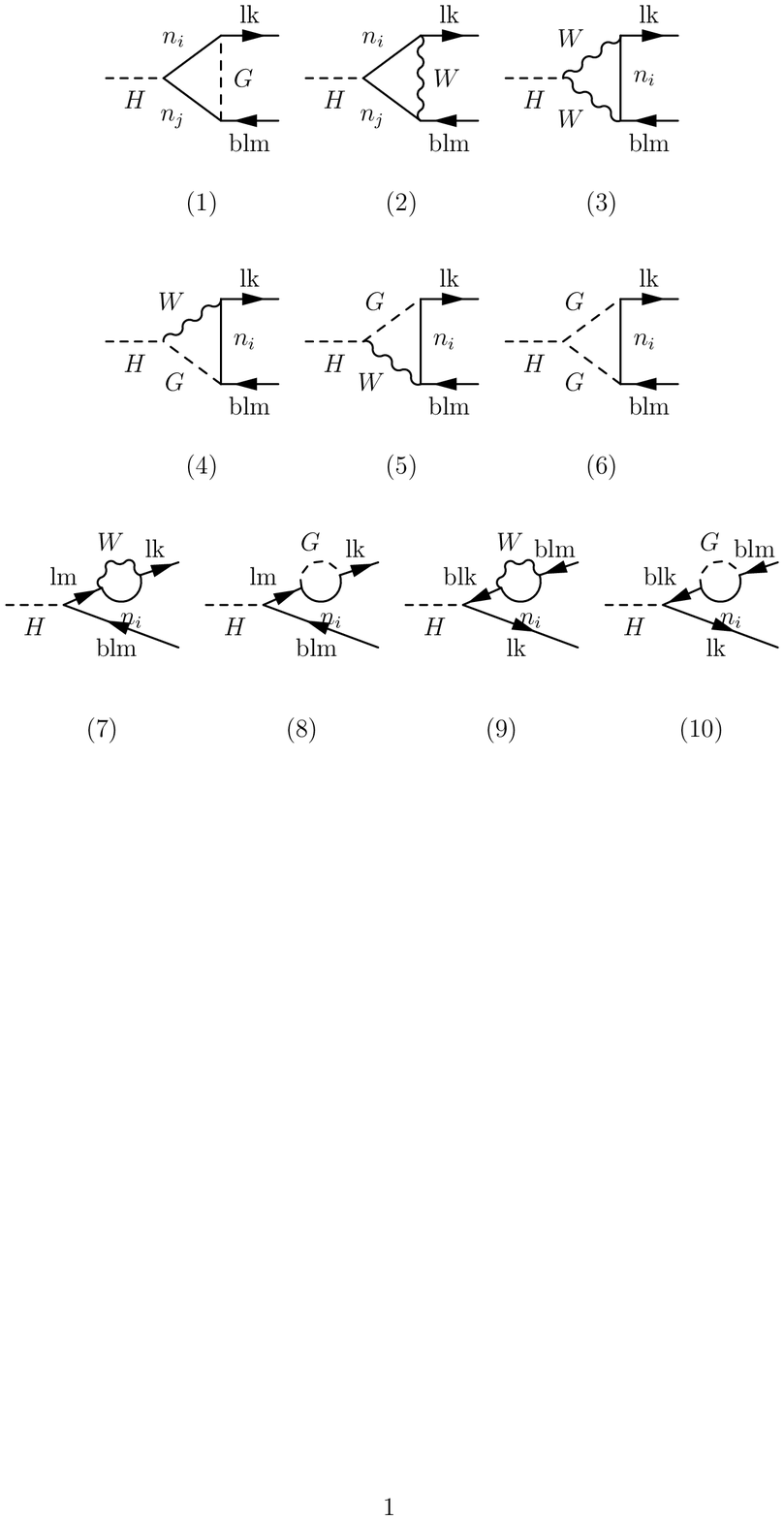}
\caption{One-loop diagrams for the LFVHD in the SM-seesaw model}
\label{fig:1}
\end{figure}

\begin{figure}
\includegraphics[width=8.0cm,height=5.5cm]{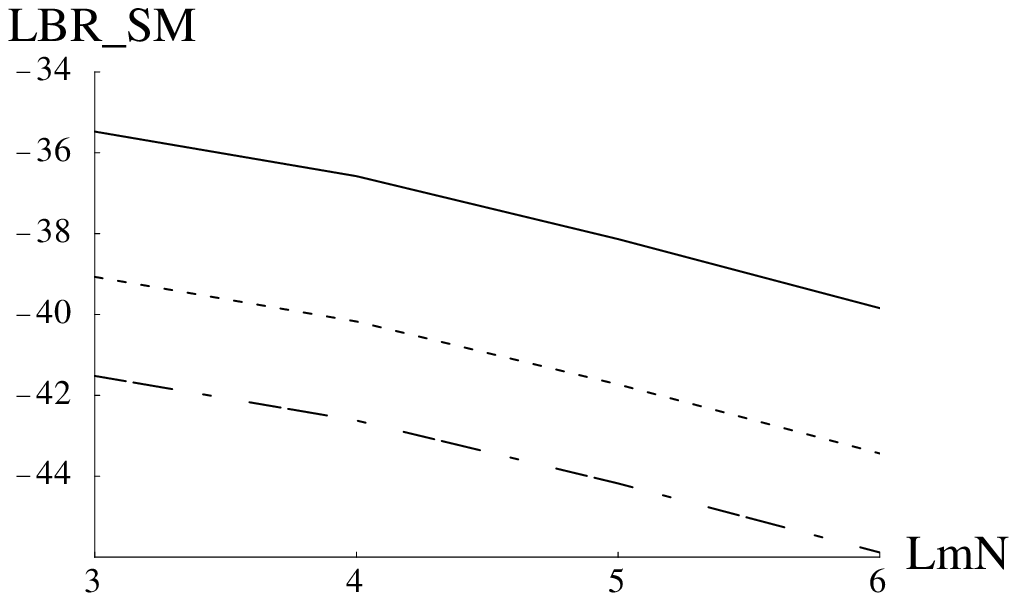}
\caption{ Dependence of the LFVHD ratios in the SM-seesaw with $m_N$ (GeV) in scenario A with degenerate heavy
  neutrinos. Solid, short-dashed and 
long-dashed lines are for $BR_{23}, BR_{13}$ and $BR_{12}$
  respectively.}
\label{fig:2}
\end{figure}

\begin{table}[h!]
\begin{center}
\begin{tabular}{|c|c|c|c|} \hline
%\multicolumn{2}{|c|}{} & \multicolumn{2}{|c|}{} \\ \hline \hline
\multicolumn{2}{|c|}{} & $R_0$ & $R_1$ \\ \hline 
(A) & $BR_{12}$ & $3 \times 10^{-42}$ & $2 \times 10^{-32}$ \\
  $m_{N_{1,2,3}} = 10^{3}$ GeV & $BR_{13}$ & $ 1 \times 10^{-39}$ &$2 \times 10^{-30}$ \\
 & $BR_{23}$ & $ 3 \times 10^{-36}$ &$3 \times 10^{-30}$ \\\hline
(A) & $BR_{12}$ & $1 \times 10^{-46}$ & $ 1 \times 10^{-36} $ \\
  $m_{N_{1,2,3}} = 10^{6}$ GeV & $BR_{13}$ & $ 3 \times 10^{-44}$ & $ 1
\times 10^{-34}$ \\
 & $BR_{23}$ & $1 \times 10^{-40}$ & $ 1 \times 10^{-34}$ \\\hline \hline
(B) & $BR_{12}$ & $1 \times 10^{-38}$ & $ 1 \times 10^{-35}$ \\
$m_{N_{1(2)}}=1 \, (5) \times 10^3$ GeV & $BR_{13}$ & $ 1 \times 10^{-
36}$ & $ 3 \times 10^{-33}$ \\
$m_{N_3}=1 \times 10^{5}$ GeV & $BR_{23}$ & $ 1 \times 10^{-37}$ & $
3 \times 10^{-32}$ \\\hline
(B) & $BR_{12}$ & $ 4 \times 10^{-39}$ & $ 1 \times 10^{-36}$ \\
$m_{N_{1(2)}}=1 \, (5) \times 10^{4}$ GeV  & $BR_{13}$ & $ 1 \times
10^{-37}$ & $ 1 \times 10^{-34}$ \\
$m_{N_3}=1 \times 10^{6}$ GeV & $BR_{23}$ & $ 1 \times 10^{-37}$ & $
2 \times 10^{-33}$ \\\hline \hline
\end{tabular}
%\vs
\caption{Branching ratios of the LFVHD in the SM-seesaw, for scenarios A and
  B and for various choices of the heavy neutrino masses. The R matrix is 
  chosen  as in case 0 and case 1 with $\theta_1=\theta_2=\theta_3=\pi/3 e^{i\pi/3}$. 
  Here, 
$BR_{12}=BR(H \to \mu {\bar e})$, $BR_{13}=BR(H \to \tau {\bar e})$, 
$BR_{23}=BR(H \to \tau {\bar \mu})$ and $m_H=115$ GeV.}
\end{center}
\label{tabla1}
\end{table} 
The main conclusion from the results in table I is that the 
LFVHD rates in
the SM-seesaw  are extremely small and depend strongly on
the Majorana mass scale. In all scenarios considered here and for 
large $m_N$, say larger than $10^4$ GeV, the decay rates decrease
with the Majorana mass, as $\sim (1/m_M^2)$, and therefore they become 
very small for very large masses of
the heavy neutrinos. 
This behaviour is easily explained by the fact that the dominant contributions 
to the form
factors are proportional to $\xi \xi^+ \sim {\mathcal O}(m_D^2/m_M^2)$ and, 
$m_D$ being obtained by eq.~\ref{Rcasas} goes as $m_D \sim m_M^{1/2}$, so
that the form factors scale as $F \sim 1/m_M$ and, consequently, 
the branching ratios as
$BR \sim 1/m_M^2$. That is, we get decoupling of the heavy neutrinos.
We have checked numerically this behaviour, as can be seen in fig~\ref{fig:2}. 
Incidentaly, it should be mentioned that since we are keeping 
the masses of the light neutrinos non-vanishing, the asymptotic value 
is not exactly zero but an extremely small value.

The R matrix that appears in the relation between $m_D$ and $m_M$
can be also relevant in some scenarios. In particular, if the $\theta_i$ angles
are complex, the R-matrix elements can have large modulus, 
the derived $m_D$ can increase relevantly and, therefore, the decay rates can 
be much larger, although still very small. 
This can be seen clearly by comparing the first and second
columns of table I. 
Notice also that, due to the leptonic mass hierarchy, the larger rates
are always obtained for the decays involving the $\tau$-lepton.

Therefore one can conclude that the LFVHD rates in the SM-seesaw 
are negligibly small
and the largest values are for the lowest Majorana mass choices and largest 
complex angles in the R matrix. These conclusions are valid for both scenarios A
and B. 

In order to compare these results with the ones obtained in~\cite{Pilaftsis1}, 
it has to be taken into account that in their computation the seesaw parameter 
 $\xi$ was fixed to a particular numerical value, so that, for large $m_M$,
 $m_D$ scales as $m_D
  \sim m_M$, in contrast to our case where, as we have said,  $m_D \sim 
  m_M^{1/2}$. These different assumptions on the relation between $m_D$
  and $m_M$ give rise to very different decay rates. The branching ratios 
   in their computation grow instead with $m_M$, thus by providing very large values
  to the Dirac masses (i.e., the Yukawa couplings get strong as $m_N \simeq m_M$ increases) they obtained  
  much larger values for the branching
  ratios. For a fixed neutrino mass $m_N$,
  the only way to reach these large values for the Dirac masses in
  our scenario is by
  considering extremely large complex R-matrix elements. As an example, 
  in scenario  A and case 1 with $m_{N_i}= 1$
  TeV and $\theta_{1, \, 2, \, 3} \sim 5i$, we obtain $|m_D| \sim 200$ GeV and
   $BR_{23} \sim
  10^{-6}$, which are closer to the predicted values in 
  ref.~\cite{Pilaftsis1}. But this is not a natural assumption in seesaw
  models since it
  requires a strong fine-tunning in the choice of the $m_D$ and $m_M$ matrices.

We also include next, for comparison, the numerical results 
for the LFVHD branching ratios in the case of Dirac neutrinos 
and for $m_H=115$ GeV. For the two scenarios A and B they are, correspondingly:
\begin{eqnarray*}
Br(H\to \mu {\bar e})&=& 2 \times 10^{-62} \, (A)\,,\,\, 2 \times 10^{-62}
\, (B)   \nonumber \\
Br(H\to \tau {\bar e})&=& 1 \times 10^{-60} \,(A)\,,\,\, 1 \times 10^{-60}
\,(B) \nonumber \\
Br(H\to \tau {\bar \mu})&=& 3 \times 10^{-56} \,(A)\,,\, 2 \times 10^{-56} \,(B)
\end{eqnarray*}
 As we can see, 
these ratios are negligible and much smaller than in the case of Majorana
neutrinos. We also learn from our results than the ratios undergo a
strong cancellation when the internal neutrinos in the loop diagrams 
are summed over 
the three generations. This 
suppression  is similar to the GIM suppression mechanism  of 
the quark sector. Notice that our rates disagree in many orders of magnitude with the results
of ref.~\cite{Diaz-Cruz:1999xe}, where a prediction of $BR(H \to \tau \bar{\mu}) \sim 6\times 10^{-7}$ was
presented. We believe that this disagreement could be due to the above mentioned 
strong cancellations
which are not taking place in their case.  

%%%%%%%%%%%%%%%%%%%%%%%%%%%%%%%%%%%%%%%%%%%%%%%%%%%%%%%%%%%%%%%%%%%%%%%%%%%%%%%%%%%%%%%
\section{\label{sec:4} Lepton Flavor Violating Higgs Decays in the MSSM-seesaw}
In this section we study the lepton flavor violating Higgs decays within the context of the 
MSSM-seesaw model. We first analyze the various sources of lepton flavor changing processes 
in this model, and next we compute the LFVHD partial widths and branching ratios for the 
three neutral MSSM Higgs boson decays, 
$\Gamma (H_x \to l_k \bar{l}_m)$ with $H_x = h_0, H_0, A_0$ and $l_k \bar{l}_m=\tau \bar{\mu}, 
\tau \bar{e}, \mu \bar{e}$. We also analyze in parallel the lepton flavor
changing $\l_j\to l_i \gamma$ decays and explore the maximum predicted rates for 
LFVHD, mainly for $H^0,A^0 \to \tau {\bar \mu}$ decays, by requiring the 
$BR(\l_j\to l_i \gamma)$
rates to be within the experimental allowed range. We use the present
experimental upper bounds given, respectively, by  
$|BR(\mu \to e \gamma)|<1.2 \times 10^{-11}$~\cite{mue}, 
$|BR(\tau \to \mu \gamma)|<3.1 \times 10^{-7}$~\cite{taumu} and 
$|BR(\tau \to e \gamma)|<2.7 \times 10^{-6}$~\cite{taue}.
%%%%%%%%%%%%%%%%%%%%%%%%%%%%%%%%%%%%%%%%%%%%%%%%%%%%%%%%%%%%%%%%%%%%%%
\subsection{Sources of LFV interactions in the MSSM-seesaw}
In the MSSM-seesaw model there are two sources of lepton flavor changing processes. 
The first one is induced from the non-vanishing mixing in the light neutrino sector, that is 
from the off-diagonal elements of the  $U_{MNS}$ matrix. We have already seen in the 
previous section
that these elements induce, via charged currents, intergenerational interactions of the type, 
$W^{\pm}-l_i-\nu_j$ (and the corresponding $G^{\pm}-l_i-\nu_j$) with $i \neq j$. These will be 
generated in the MSSM-seesaw as well, but in addition, there will appear new tree-level  LFV 
interactions 
involving neutrinos, as for instance those with  
charginos $\tilde{\chi}^{\pm}$ and charged sleptons $\tilde{l}$, 
 $\tilde{\chi}^{\pm}-\tilde{l}_i-\nu_j$  with 
$i \neq j$, and those with neutralinos $\tilde{\chi}^{0}$ and sneutrinos $\tilde{\nu}$, 
$\tilde{\chi}^{0}-\tilde{\nu}_i-\nu_j$  with 
$i \neq j$. From the
operational point of view, these effects are introduced by the explicit factors of 
the $U_{MNS}$ 
matrix elements in the corresponding couplings when referred to the physical
basis. 
    Due to the extended Higgs sector of the MSSM-seesaw, there will be 
   additional 
 intergenerational couplings of the type $H^{\pm}l_i\nu_j$ 
   with similar $(U_{MNS})_{ij}$ factors. 

All the previous intergenerational couplings can induce 
important contributions to LFV processes with external SM leptons. In the following we will refer to the
radiatively induced flavor changing effects from the $(U_{MNS})_{ij}$ factors in
the couplings as $U_{MNS}$ effects.  For the case under study here of LFVHD, 
the couplings inducing these $U_{MNS}$ effects are just $W^{\pm}l_i\nu_j$, $G^{\pm}l_i\nu_j$ and 
$H^{\pm}l_i\nu_j$.  
The size of the generated effects from the $W^{\pm}l_i\nu_j$ and $G^{\pm}l_i\nu_j$ couplings 
are expected to be as 
small as in the SM-seesaw case. The only difference comes from the different couplings of 
the internal particles to the external $h_0, H_0$, and $A_0$ Higgs bosons as compared 
to those of the SM Higgs boson H, but this difference will not change relevantly the small 
size of the generated ratios.
On the other hand, the generated effects from the $H^{\pm}l_i\nu_j$ couplings are new with respect 
to the SM-seesaw case. We have numerically estimated the size of 
the contributions to the LFVHD branching ratios from all the one-loop diagrams with internal 
$H^{\pm}$, $W^{\pm}$ and $G^{\pm}$, and we have found that they are indeed always negligibly small.  
We will in consequence ignore all these contributions in the following.  

The second source of LFV processes in the MSSM-seesaw is genuine in SUSY 
models and comes from the misalignment between the rotations leading to the mass eigenstate basis 
of sleptons relatively to that of leptons. This misalignment is radiatively generated 
in the SUSY-seesaw models from the Yukawa couplings of the Majorana neutrinos and 
can be sizeable in both the charged slepton and sneutrino sectors . 
Once one rotates the so-generated flavor non-diagonal charged slepton and sneutrino 
mass matrices to the physical diagonal 
ones,  some intergenerational couplings involving SUSY particles are generated. For the 
 case of LFVHD, the involved couplings are 
$\tilde{\chi}^{\pm}l_i\tilde{\nu}_j$ and $\tilde{\chi}^{0}l_i\tilde{l}_j$ 
with $i \neq j$. These   generated effects from lepton-slepton misalignment will be called 
in the following misalignment effects. 
All these couplings will induce via loops of SUSY particles relevant contributions 
 to the LFVHD rates as will
 be shown later on. 

The LFV effects from misalignment are usually implemented in the seesaw models in the 
language of the Renormalization Group Equations (RGE) and can be summarized as follows. 
One starts with universal soft-SUSY-breaking parameters at the large energies 
$M_X >> m_M$ given by,
\begin{eqnarray}
(m_{\tilde{L}})_{ij}^2 = M_0^2 \delta_{ij}, \, 
(m_{\tilde{E}})_{ij}^2 = M_0^2 \delta_{ij}, \, 
(m_{\tilde{M}})_{ij}^2 = M_0^2 \delta_{ij},
\, (A_{l})_{ij}= A_0 (Y_l)_{ij}, \, (A_{\nu})_{ij}= A_0 (Y_{\nu})_{ij}
\label{univ_cond}
\end{eqnarray}
where, $M_0$ and $A_0$ are the universal soft-slepton mass and soft-trilinear 
coupling respectively.
 $Y_l$ is the Yukawa coupling matrix of the charged leptons, which is flavor 
 diagonal in the basis chosen here, $(Y_{l})_{ij}=Y_{l_i} \delta_{ij}$ with 
 $Y_{l_i}= \frac{m_{l_i}}{v_1}$ and $v_1=v \cos \beta$. $Y_{\nu}$ is the Yukawa coupling
 matrix of the neutrinos and is given by, 
 $(Y_{\nu})_{ij}=\frac{(m_D)_{ij}}{v_2}$, with $v_2=v\sin\beta$.
 
The effect of the RGE-running from $M_X$ down to $m_M$ on the off-diagonal soft parameters 
of the charged slepton sector,  to one-loop and in the leading-log approximation, 
is then described by, 
\begin{eqnarray}
(\Delta m_{\tilde{L}}^2)_{ij}&=&-\frac{1}{8 \pi^2} (3 M_0^2+ A_0^2) (Y_{\nu}^* L Y_{\nu}^T)_{ij} \nonumber \\
(\Delta A_l)_{ij}&=&- \frac{3}{16 \pi^2} A_0 Y_{l_i} (Y_{\nu}^* L Y_{\nu}^T)_{ij}\nonumber \\
(\Delta m_{\tilde{E}}^2)_{ij}&=&0\,\,;\, L_{kl} \equiv \log \left( \frac{M_X}{m_{M_k}}\right) \delta_{kl}.
\end{eqnarray}
Notice that our notation is slightly different from the usual one in that 
$Y_{\nu} \longleftrightarrow Y_{\nu}^T$. 
For all the numerical estimates in this work we use $M_X=2 \times 10^{16}$ GeV and, for simplicity, we will 
assume $A_0=0$. Thus, we only consider flavor changing in the LL sector, 
which is known to be a very good
approximation within the context of the seesaw model. In fact, we have checked that even for values as large as 
$A_0=M_0=1 TeV$ the size of the corresponding flavor changing dimensionless parameters in the LR sector
are always smaller than the LL ones in more than three orders of magnitude. 
Notice also that, in addition
to the previously mentioned  $U_{MNS}$ factors, 
which, as we have said, are not relevant for the present computation of the LFVHD rates, 
there is an extra 
dependence on $U_{MNS}$ via the $Y_{\nu}$ couplings or, equivalently, via the $m_D$ matrix as given in 
eq.~\ref{Rcasas}. Thus, even if we fixed $U_{MNS}$ to the unit matrix, there could be still flavor changing effects
from misalignment via the $R$ matrix. Conversely, if we fixed $R$ to the unit matrix, 
there could be still
flavor changing effects from misalignment via 
the $U_{MNS}$ matrix in $m_D$.  Finally, notice that   
the effect of neutrino Yukawa couplings, via  RGE-running from $M_X$ down to $m_M$, on the 
diagonal entries of the squared-mass
matrices are small, and have been neglected  in this work. On the other hand, as we have
previously said, we have included the effects of the RGE-running  from $m_W$ up to $m_N$ on
the neutrino masses $m_{\nu_i}$ $(i=1,2,3)$ and the $U_{MNS}$ matrix elements, and we have
used the corresponding corrected parameters in eq.~\ref{Rcasas} to evaluate the Yukawa
couplings $Y_{\nu}$.

\begin{figure}
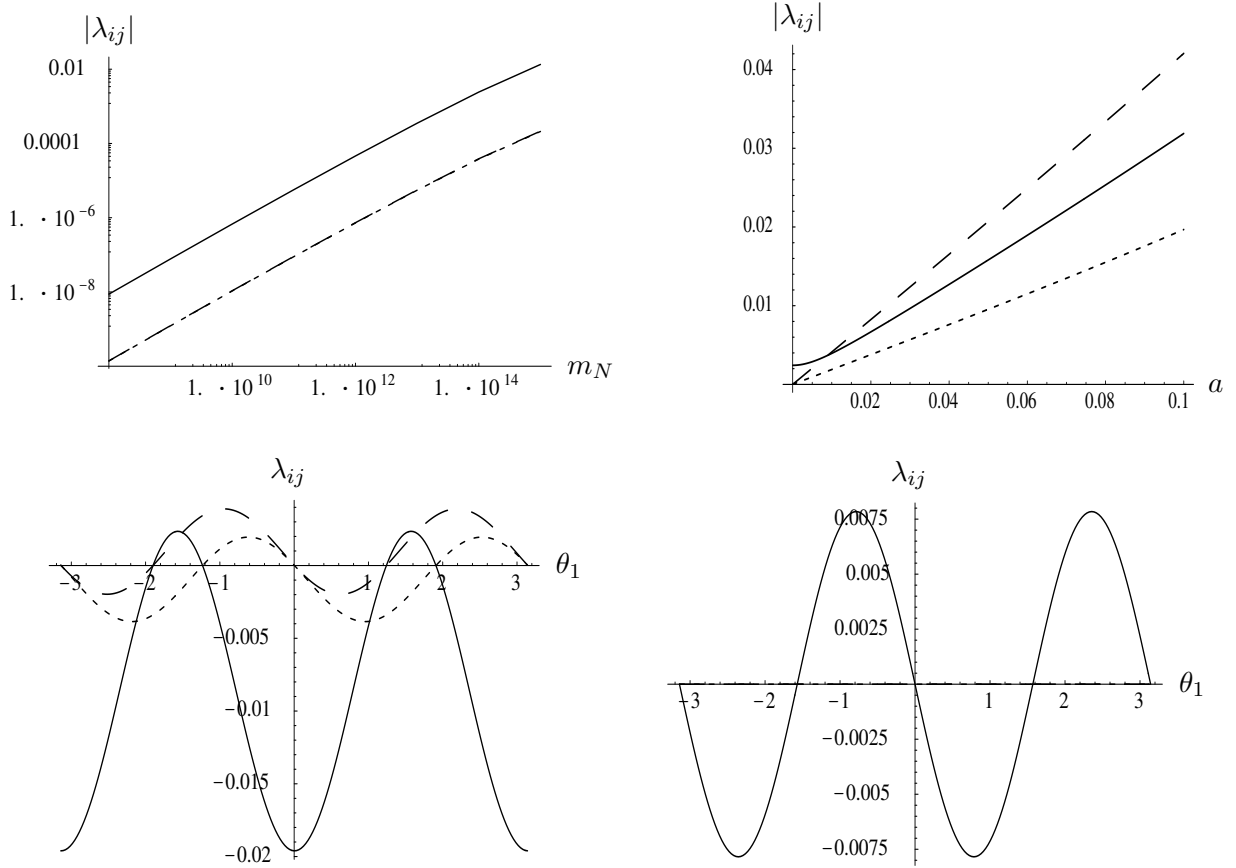

\includegraphics[width=8.0cm,height=5.5cm]{fig3a.epsi}
\hspace{1.5cm}
\includegraphics[width=6.5cm,height=5.5cm]{fig3b.epsi}

\vspace{0.5cm}
\hspace{0.5cm}
\includegraphics[width=7.5cm,height=5.5cm]{fig3c.epsi}
\hspace{0.5cm}
\includegraphics[width=7.5cm,height=5.5cm]{fig3d.epsi}
\caption{ Dependence of $\lambda_{ij}$  with the seesaw parameters. Solid, long-dashed and 
short-dashed lines are for $\lambda_{23}, \lambda_{13}$ and $\lambda_{12}$
respectively. {\bf (3a)} 
 Upper left panel: Dependence
 with $m_N$ in scenario A for real R. {\bf (3b)} Upper right panel: Dependence with the $a$ parameter 
 in scenario A with $m_N=10^{14}$ GeV and complex R, for
 case 2 with $a=b=c$. {\bf (3c)} Lower left panel: Dependence with real $\theta_1$ in scenario B, case 1, for 
$(m_{N_1},m_{N_2},m_{N_3})=(10^8, 2 \times 10^8, 10^{14})$ GeV and
$\theta_2=\theta_3=0$.{\bf (3d)} Lower right panel: Same as in (3c) but for 
$U_{MNS}=1$. In all plots here, $\tan\beta=35$.}
\label{fig:3}
\end{figure}

\begin{figure}
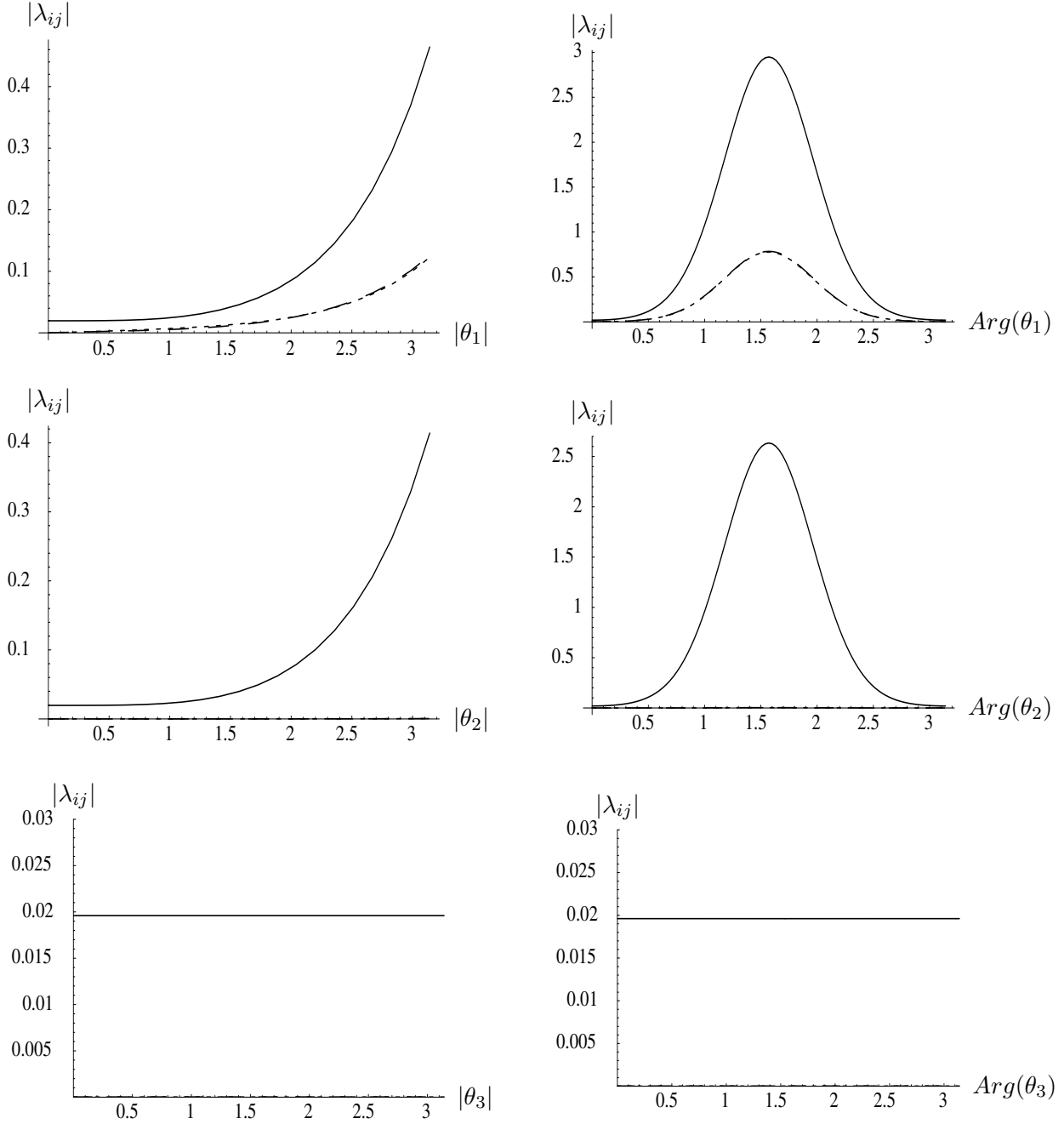

\hspace{-0.5cm}
\includegraphics[width=8.0cm,height=6.0cm]{fig4a.epsi}
\hspace{0.25cm}
\includegraphics[width=8.0cm,height=6.0cm]{fig4b.epsi}
\hspace{0.25cm}
%\vspace{0.5cm}
\\
\hspace{-0.5cm}
\includegraphics[width=8.0cm,height=6.0cm]{fig4c.epsi}
\hspace{0.25cm}
\includegraphics[width=8.0cm,height=6.0cm]{fig4d.epsi}
\hspace{0.25cm}
%\vspace{0.5cm}
\\
\hspace{-0.5cm}
\includegraphics[width=8.0cm,height=6.0cm]{fig4e.epsi}
\hspace{0.25cm}
\includegraphics[width=8.0cm,height=6.0cm]{fig4f.epsi}
\caption{ Dependence of $|\lambda_{ij}|$  with the seesaw parameters for complex
$\theta_i$. Solid, long-dashed and 
short-dashed lines (the two later being undistinguishible in the plots) 
are for $|\lambda_{23}|$, $|\lambda_{13}|$ and $|\lambda_{12}|$ respectively.
Left panels: Dependence
 with $|\theta_1|$, $|\theta_2|$ and $|\theta_3|$ respectively in scenario B, case 1. 
 We take $Arg(\theta_i)=\pi/4$, correspondingly, and 
 the rest of angles are set to zero. Right panels: 
 Dependence with $Arg(\theta_1)$, $Arg(\theta_2)$ and $Arg(\theta_3)$ respectively in 
 scenario B, case 1. We take $|\theta_i|=\pi$, correspondingly, and the rest of angles are set to 
 zero. In all plots here, $(m_{N_1},m_{N_2},m_{N_3})=(10^8, 2 \times 10^8, 10^{14})$ GeV
 and  $\tan\beta=35$.}
\label{fig:4}
\end{figure}

In order to estimate the size of the misalignment effects,
we first study in detail the dependence of the flavor changing dimensionless parameters, in the LL
sector, defined as
 $\lambda_{ij} \equiv \frac{(\Delta m_{\tilde{L}}^2)_{ij}}{M_0^2}$, as a function of
 the seesaw parameters. We show in figs.~(\ref{fig:3}) and ~(\ref{fig:4}) the predictions of the $\lambda_{ij}$ parameter as a function of
 the seesaw parameters, and in some selected examples within the scenarios 
described above. We show in fig.~(\ref{fig:3}a) the dependence of $|\lambda_{ij}|$ with $m_N$, for 
 scenario A and real R. 
 Notice that in this case, the $\lambda_{ij}$ turn out to be independent on R, so this 
figure applies 
both to case 0 and to case 1 with $\theta_i$ real. As can be seen, the three $|\lambda_{ij}|$ 
grow with $m_N$ and the largest one, which is  
$|\lambda_{23}|$, reaches values up to  $2.4 \times 10^{-3}$ for $m_N=10^{14}$ GeV. 
The size of $|\lambda_{12}|$ can reach values  up to $3.8 \times 10^{-5}$ for 
$m_N=10^{14}$ GeV, and correspondingly for $|\lambda_{13}| \simeq |\lambda_{12}|$. The size of the $\lambda_{ij}$
in this scenario with degenerate heavy neutrinos can obviously be increased if R is assumed 
instead to be complex. In this case, the $\lambda_{ij}$
 are in general complex numbers. As an example, we show in fig.~(\ref{fig:3}b) the dependence of $|\lambda_{ij}|$, 
 in scenario A with $m_N= 10^{14}$ GeV, on the parameter $a$ of case 2, for $a=b=c$ . The size of 
$|\lambda_{ij}|$ increases clearly with $a$ and, for the studied range, $|\lambda_{12}|$ can be 
as large as $2 \times 10^{-2}$. In particular, for values of $|abc|\simeq 10^{-5}$, that 
generate succesfull baryogenesis~\cite{Pascoli:2003rq}, the generated 
$|\lambda_{ij}|$ are still large, namely, $|\lambda_{12}|=4\times 10^{-3}$.
Notice also 
that the relative size of the different   $\lambda_{ij}$ changes with
respect to fig.~(\ref{fig:3}a). The main problem with this scenario A
and R complex, with the parametrization of case 2,  is
that for most choices of the relevant parameters, $m_{N}$, $a$, $b$ and $c$,
producing 
succesfull baryogenesis, the size of the generated $\lambda_{12}$ is too large and
leads to large $\mu \to e \gamma$ decay rates~\cite{Pascoli:2003rq} clearly above the present experimental bound
of $BR(\mu \to e \gamma) < 1.2 \times 10^{-11}$~\cite{mue}. We have also studied the
scenario A with complex R given by the parametrization of case 1, and have found  
that 
for most choices of the $\theta_i$ complex angles the three $\lambda_{12}$,
$\lambda_{13}$ and $\lambda_{23}$ have comparable sizes and therefore, the required 
large $\lambda_{23}$ values leading to large $H \to \tau {\bar \mu}$ ratios imply again
experimentally unallowed $\mu \to e \gamma$ ratios.    
 Therefore, we will not consider this scenario A with complex R anymore in the following.

The case of hierarchical neutrinos is shown in figs.~(\ref{fig:3}c) and ~(\ref{fig:3}d)
for real R and in fig.~(\ref{fig:4}) for complex R with the parametrization of case 1.
In fig.~(\ref{fig:3}c) it is plotted the dependence with real 
$\theta_1$ for scenario B with 
$(m_{N_1},m_{N_2},m_{N_3})= (10^8,2\times 10^8,10^{14})$ GeV and 
$\theta_2=\theta_3=0$. We see that $\lambda_{23}$ is the 
largest one and 
reaches negative values  up to $-1.96 \times 10^{-2}$ at the points $\theta_1=0,\,\pm \pi$, precisely where 
$\lambda_{12}$ and $\lambda_{13}$ vanish. These points are therefore particularly 
interesting for the LFVHD rates since they will lead to the largest ratios for $H \to \tau
\bar{\mu}$ while keeping $\mu \to e \gamma$, $\tau \to e \gamma$ extremely small and, as we
will see later, 
$\tau \to \mu \gamma$ still compatible with data. 
Notice that the
point $\theta_1=0$ in  fig.~(\ref{fig:3}c) corresponds to the reference case 0 
with $R=1$ and, therefore, represents the situation where the $U_{MNS}$ matrix is the only
origin for flavor changing. This means that our experimental input for the $U_{MNS}$ matrix
generates by itself sizeable rates for lepton flavor violating decays involving the second
and third generations. The alternative situation where just the $R$ matrix is generating the 
lepton flavor violating decays is illustrated in fig.~(\ref{fig:3}d). Here we show the 
$\lambda_{ij}$ dependence with $\theta_1$ for $\theta_2=\theta_3=0$ and $U_{MNS}=1$. We see
that $|\lambda_{23}|$ reaches values up to about $7.8 \times 10^{-3}$ whereas 
$\lambda_{12}$ and $\lambda_{13}$ vanish for all $\theta_1$. By comparing these two
situations we can infere that, for the case of real R with $\theta_1 \neq 0$
 and $\theta_2=\theta_3=0$, the 
induced misalignment effect between the second and third generations from the experimental
$U_{MNS}$ is relevant and can be even larger than the effect from R. 

We have also studied the alternative choices for real R  with $\theta_2 \neq 0$, 
$\theta_1=\theta_3=0$  and 
with $\theta_3 \neq 0$, $\theta_1=\theta_2=0$, although the corresponding plots are not 
shown here for
brevity. We find a $\lambda_{23}$ dependence on $\theta_2$ very similar to that on
$\theta_1$, with maximum negative $\lambda_{23}$ values at $\theta_2=0, \pm \pi$ of $-1.96 \times 10^{-2}$. In
contrast, $\lambda_{12}$ and $\lambda_{13}$ now take 
very small values whose maximum are $2.4 \times 10^{-5}$.  Regarding the dependence 
with $\theta_3$  a different situation is found, where the three $\lambda_{12}$, 
$\lambda_{13}$ and 
 $\lambda_{23}$ are approximately constant with $\theta_3$ and take the values, 
 $\lambda_{23}=-1.96 \times 10^{-2}$, and 
 $\lambda_{13}\simeq \lambda_{12}=2.4 \times 10^{-5}$, respectively. 
 
 The case of hierarchical neutrinos with complex R  produces, in most cases, 
 complex $\lambda_{ij}$ values and their moduli are in general larger than in the case
 of real R, as can be clearly seen in fig.~(\ref{fig:4}). Our particular choice 
 for the heavy neutrino masses of $(m_{N_1},m_{N_2},m_{N_3})= (10^8,2\times 10^8,10^{14})$ GeV,
 where
 the two lightest neutrinos have similar masses and well below the mass of the heaviest one, 
 produces the
 specific pattern shown in these plots, where the dependence of $|\lambda_{23}|$ on $\theta_1$,
 for $\theta_2=\theta_3=0$, and on $\theta_2$, for $\theta_1=\theta_3=0$ are very similar, 
 and $|\lambda_{23}|$ can reach very large values for a large region  of the 
 $(|\theta_i|$,$Arg(\theta_i))$, $i=1,2$ parameter space. For instance, for fixed
 $Arg(\theta_1)=\pi/4$, and $|\theta_1|$ up to $\pi$ we find $|\lambda_{23}|$ values up to 
 $0.46$ and similarly for $\theta_2$. Larger values of $Arg(\theta_i)$, $i=1,2$, 
 produce even larger $|\lambda_{23}|$ and it reaches its maximum at  
 $Arg(\theta_i)=\pi/2$. 
 In contrast, $\lambda_{12}$ and $\lambda_{13}$ reach much smaller values with complex
 $\theta_2$ than with complex $\theta_1$, being  
 $|\lambda_{12}|\simeq |\lambda_{13}| < 5 \times 10^{-4}$ for $|\theta_2|<\pi$.
  On the other hand, they depend strongly with 
 complex $\theta_1$ and $|\lambda_{12}|$ can reach too large values, up to ${\cal O}(10^{-1})$, 
 in clear conflict
 with the allowed values by $\mu \to e \gamma$ data. Finally, the behaviour with
 complex $\theta_3$ is very similar to the real case, with the three $|\lambda_{12}|$, 
 $|\lambda_{13}|$ and $|\lambda_{23}|$ being nearly constant with $\theta_3$.  Their values are $|\lambda_{23}|=1.96 \times 10^{-2}$ and 
 $|\lambda_{12}| \simeq |\lambda_{13}| =2.4 \times 10^{-5}$, respectively.
     
 Regarding the values of the hierarchical heavy neutrino masses, we have also tried 
 other choices in the range $10^{8} GeV  \leq  m_{N_i} \leq 10^{14} GeV$ and found 
 that, under the favored assumption by baryogenesis of close $m_{N_1}$ and $m_{N_2}$ 
 and much lighter than $m_{N_3}$, it is the value of this later what matters for 
 $BR(H \to \tau {\bar \mu})$, leading to
 larger $\lambda_{23}$ values for larger $m_{N_3}$ and, therefore, 
 our choice of $(m_{N_1},m_{N_2},m_{N_3})= (10^8,2\times 10^8,10^{14})$ GeV 
 seems to be appropriate. We have checked that alternative choices where $m_{N_2}$ and $m_{N_3}$ are
 close and much heavier than $m_{N_1}$ lead to a similar situation for $\theta_1 \neq 0$,
 $\theta_2=\theta_3=0$ than the previous case, but it gives larger preditions for
 $\lambda_{12}$ in conflict with $\mu \to e \gamma$ data. For the following 
 predictions of decay
 rates in the case of hierarchical neutrinos we will fix the values to 
 $(m_{N_1},m_{N_2},m_{N_3})= (10^8,2\times 10^8,10^{14})$ GeV.
 
 In summary, the case of hierarchical heavy neutrinos leads to 
 larger $\lambda_{ij}$ values 
 than the degenerate case and, in consequence, larger  LFVHD rates. 
 Futhermore,
  in order to get the largest possible 
 $BR(H \to \tau {\bar \mu})$ rates while keeping all $BR(l_j \to l_i \gamma)$
 rates within the experimental allowed regions, the most favorable case of
 all the studied ones, and for our choice of neutrino masses, is the one with complex 
 $\theta_2 \neq 0$, and $\theta_1=\theta_3=0$.

The next stept is the diagonalization of the charged slepton and sneutrino mass matrices 
leading to the mass eigenvalues and mass eigenstates at the electroweak energy scale.
We start with the non-diagonal charged slepton and sneutrino squared-mass 
matrices that
are obtained  after the running from $M_X$ to $M_W$ and 
 once the charged leptons and neutrinos have been rotated to the 
 physical basis.
 For the charged sector, this matrix is referred to the 
$(\tilde{e}_L, \tilde{e}_R, \tilde{\mu}_L, \tilde{\mu}_R, \tilde{\tau}_L, \tilde{\tau}_R)$ basis  and  
can be written as follows,

\begin{equation}
M_{\tilde{l}}^2\ =\ \left( \barr{cccccc} M_{LL}^{ee \, 2} & M_{LR}^{ee \, 2} & M_{LL}^{e \mu \, 2} & 0 & M_{LL}^{e \tau \, 2} & 0 \\ M_{RL}^{ee \, 2} & M_{RR}^{ee \, 2} & 0 & 0 & 0 & 0 \\ M_{LL}^{\mu e \, 2} & 0 & M_{LL}^{\mu \mu \, 2} & M_{LR}^{\mu \mu \, 2} & M_{LL}^{\mu \tau \, 2} & 0 \\ 0 & 0 & M_{RL}^{\mu \mu \, 2} & M_{RR}^{\mu \mu \, 2} & 0 & 0 \\ M_{LL}^{\tau e \, 2} & 0 & M_{LL}^{\tau \mu \, 2} & 0 & M_{LL}^{\tau \tau \, 2} & M_{LR}^{\tau \tau \, 2}\\0 & 0 & 0 & 0 & M_{RL}^{\tau \tau \, 2} & M_{RR}^{\tau \tau \, 2} \earr \right)
\end{equation}
where,
\begin{eqnarray}
M_{LL}^{ll \, 2}&=& m_{\tilde{L}, l}^2 + m_l^2 + m_Z^2 \cos 2 \beta
(-\frac{1}{2}+ \sin^2 \theta_{W}) \nonumber \\
M_{RR}^{ll \, 2}&=& m_{\tilde{E}, l}^2 + m_l^2 - m_Z^2 \cos 2 \beta \sin^2
\theta_{W} \nonumber \\
M_{LR}^{ll \, 2}&=& M_{RL}^{ll \, 2}=m_l (A_l -\mu \tan \beta) \nonumber \\
M_{LL}^{e \mu \, 2}&=& (\Delta m_{\tilde{L}}^2)_{12}
\,;\, M_{LL}^{\mu e \, 2} = (\Delta m_{\tilde{L}}^2)_{21} \nonumber \\
M_{LL}^{e \tau \, 2}&=& (\Delta m_{\tilde{L}}^2)_{13} 
\,;\,M_{LL}^{\tau e \, 2} = (\Delta m_{\tilde{L}}^2)_{31}\nonumber \\
M_{LL}^{\mu \tau \, 2}&=& (\Delta m_{\tilde{L}}^2)_{23} 
\,;\,M_{LL}^{\tau \mu \, 2} = (\Delta m_{\tilde{L}}^2)_{32}.\nonumber \\
\end {eqnarray}
 The soft SUSY breaking masses and trilinear couplings above, 
 $m_{\tilde{L}, e}$, $m_{\tilde{L}, \mu}$, $m_{\tilde{L}, \tau}$ and $A_l$, refer to their 
 corresponding values at the electroweak scale. We got them by solving
 numerically the RGE with the program mSUSPECT~\cite{Djouadi:2002ze} and by imposing the 
 universality conditions  at $M_X=2\times 10^{16} $ GeV for the sfermion sector, eq.~\ref{univ_cond},  together with the corresponding ones for 
 the gaugino and Higgs boson sectors, $M_1(M_X)=M_2(M_X)=M_{1/2}$ and 
 $M_{H_1}(M_X)=M_{H_2}(M_X)=M_0$ respectively. For the gaugino sector, this implies 
 the well known relation at low energies, $M_1(M_W)=\frac{5}{3}(\tan\theta_W)^2M_2(M_W)$.
 The value of the supersymmetric 
 $\mu$ parameter is extracted 
 as usual from the electroweak breaking condition. We choose in all this paper, $\mu >0$, and do not 
 expect relevant differences for $\mu<0$.   
 
After diagonalization of the $M_{\tilde{l}}^2$ matrix one gets the physical 
slepton masses and the six mass eigenstates ($\tilde{l_1},.....,\tilde{l_6}$)$\equiv \tilde{l}$ 
which are related to the previous weak eigenstates 
($\tilde{e}_L$,....$\tilde{\tau}_R$)$\equiv \tilde{l}'$ by the corresponding $6 \times 6$ rotation 
matrix, $\tilde{l}' = R^{(l)}\tilde{l}$. 

Regarding the sneutrino sector one proceeds similarly to the charged slepton sector, 
but now the diagonalization process is simpler because of the involved seesaw matrix 
$\xi$ which gives rise naturally to a suppresion of the RH sneutrino components in the relevant 
mass eigenstates. This can be easily illustrated in the one generation case, 
but for three generations one arises to similar conclusions. 
The sneutrino mass terms of the MSSM-seesaw model can be written in the one 
generation case~\cite{Grossman:1997is} as,

\begin{equation}
-\mathcal{L}_{mass}^{\nu}=\left(\barr{c} Re (\tilde{\nu}_L) \, Re (\tilde{\nu}_R) \, 
Im(\tilde{\nu}_L) \, Im (\tilde{\nu}_R)\earr \right)
\left( \barr{cc} M_+^2 & 0\\ 0 & M_{-}^2 \earr \right)
 \left(\barr{c} Re(\tilde{\nu}_L) \\  Re(\tilde{\nu}_R) \\ Im(\tilde{\nu}_L)\\ Im(\tilde{\nu}_R) \earr \right)
\end{equation}

with,
\begin{equation}
M_{\pm}^2=
\left( \barr{cc} m_{\tilde{L}}^2+ m_D-\frac{1}{2}m_Z^2 \cos 2 \beta & 
m_D (A_{\nu}- \mu \cot \beta \pm m_M)\\  m_D (A_{\nu}- \mu \cot \beta \pm m_M) & m_{\tilde{M}}^2+m_D^2+m_M^2 \pm 2 B_M m_M \earr \right)
\end{equation}

Notice that now there are several mass scales involved, the soft SUSY-breaking parameters, 
$m_{\tilde L}$, $m_{\tilde M}$, $B_M$ and $A_{\nu}$, 
the Dirac mass $m_D$, the $\mu$-mass parameter, the Z boson mass $m_Z$ and the Majorana 
neutrino mass $m_M$. Our basic assumption in all this paper is that $m_M$ is much heavier 
than the other mass scales involved (obviously, except $M_X$), 
$m_M>>m_D, m_Z, \mu, m_{\tilde{L}}, m_{\tilde M}, A_{\nu}, B_M$. The size of $B_M$ has been discussed in the 
literature~\cite{Grossman:1997is} and seems more controversial. For simplicity, we shall assume here this is also 
smaller than $m_M$. In this situation,  
the diagonalization of the previous sneutrino squared mass matrix is simpler and leads to 
four mass eigenstates, two of which are light, $\xi_1^l$, $\xi_2^l$ and two heavy, 
$\xi_1^h$, $\xi_2^h$. In the leading orders of the series expansion in powers of $\xi$ the mass 
eigenstates and their corresponding mass eigenvalues are given by,

\begin{eqnarray}
\xi_1^l &=& \sqrt{2} \left( Re(\tilde{\nu}_L)- \xi Re(\tilde{\nu}_R)\right) \,\,;
\xi_2^l = \sqrt{2} \left( Im(\tilde{\nu}_L)- \xi Im(\tilde{\nu}_R)\right) \nonumber \\
\xi_1^h &=& \sqrt{2} \left( Re(\tilde{\nu}_R)+ \xi Re(\tilde{\nu}_L)\right) \,\,;
\xi_2^h = \sqrt{2} \left( Im(\tilde{\nu}_R)- \xi Im(\tilde{\nu}_L)\right) \nonumber \\
m_{\xi_{1,2}^l}^2 &=& m_{\tilde{L}}^2+ 
\frac{1}{2} m_Z^2 \cos 2 \beta \mp 2 m_D (A_{\nu}-\mu \cot \beta-B_N)\xi \nonumber \\
m_{\xi_{1,2}^h}^2 &=& m_M^{2} \pm 2 B_M m_M +m_{\tilde M}^2 + 2 m_D^2
\end{eqnarray}  

Here we can see that the heavy states $\xi_{1,2}^h$ will couple 
very weakly to the rest of particles of the MSSM via their $\tilde{\nu}_L$ component, 
which is highly suppresed by the small factor $\xi$ and, therefore, it is a good approximation to 
ignore them and keep just the light states $\xi_{1,2}^l$, which are made mainly of $\tilde{\nu}_L$ 
and its complex conjugate $\tilde{\nu}_L^*$. Now, by working in this simplified basis but applied 
to the three generations case, which we write for short $\tilde{\nu}_{\alpha}'$ ($\alpha = 1,2,3$), 
the relevant $3 \times 3$ sneutrino squared mass matrix can be written as follows,

\begin{equation}
M_{\tilde{\nu}}^2\ =\ \left( \barr{ccc} m_{\tilde{L}, e}^2 + \frac{1}{2} m_Z^2 \cos 2 \beta  & (\Delta m_{\tilde{L}}^2)_{12} & (\Delta m_{\tilde{L}}^2)_{13}\\(\Delta m_{\tilde{L}}^2)_{21}  &  m_{\tilde{L}, \mu}^2 + \frac{1}{2} m_Z^2 \cos 2 \beta & (\Delta m_{\tilde{L}}^2)_{23} \\ (\Delta m_{\tilde{L}}^2)_{31} & (\Delta m_{\tilde{L}}^2)_{32} &  m_{\tilde{L}, \tau}^2 + \frac{1}{2} m_Z^2 \cos 2 \beta  \earr \right)
\end{equation}
where $m_{\tilde{L}, l}^2$ and $(\Delta m_{\tilde{L}}^2)_{ij}$ are the same as
in the previous charged slepton squared mass matrix. 
After diagonalization of the $M_{\tilde{\nu}}^2$ matrix one gets the relevant physical sneutrino masses and eigenstates, $\tilde{\nu}_{\beta}$ ($\beta = 1,2,3$) which are related to the previous states $\tilde{\nu}_{\alpha}'$ by the corresponding $3 \times 3$ rotation matrix, $\tilde{\nu}'=R^{(\nu)} \tilde{\nu}$.

To end up this subsection, we summarize all the interaction terms that are relevant for the computation 
of the LFVHD rates. 
We present these interactions in the physical mass eigenstate basis and will perform
all the computations in this basis. It implies   
 diagonalization in all the involved SUSY sectors, charged sleptons, sneutrinos,
 charginos,  neutralinos and Higgs bosons. 
The SUSY-electroweak interaction terms among charginos, leptons and sneutrinos 
and among neutralinos, leptons and sleptons, that are the responsible for the 
LFVHD are as follows, 
\begin{eqnarray}
\mathcal{L}_{\tilde \chi_j^- l \tilde \nu_{\alpha} } &=& 
-g\, \bar{l} \left[ A_{L \alpha j}^{(l)} P_L + 
A_{R \alpha j}^{(l)} P_R \right] \tilde \chi_j^- \tilde \nu_{\alpha} + h.c., \nonumber \\
\mathcal{L}_{\tilde \chi_a^0 l \tilde l_{\alpha} } &=& 
-g\bar{l} \left[ B_{L \alpha a}^{(l)} P_L + 
B_{R \alpha a}^{(l)} P_R \right] \tilde \chi_a^0 \tilde l_{\alpha} + h.c.
\end{eqnarray} 
where the coupling factors $A_{L \alpha j}^{(l)}$, $A_{R \alpha j}^{(l)}$, $B_{L \alpha a}^{(l)}$,
and $B_{R \alpha a}^{(l)}$ are given in  Appendix B. 

The other interaction terms that enter in the computation of the LFVHD rates are the Higgs--lepton--lepton, Higgs--sneutrino--sneutrino, Higgs--slepton--slepton, Higgs--chargino--chargino,
and Higgs--neutral\-ino--neutral\-ino interactions, reading:
\begin{eqnarray}
\mathcal{L}_{H_xll} &=& 
-g H_x\bar{l} \left[ S_{L,l}^{(x)} P_L + S_{R,l}^{(x)} P_R \right]  l \,\,,\nonumber\\ 
\mathcal{L}_{H_x \tilde s_{\alpha} \tilde s_{\beta}}&=&
-iH_x \left[ g_{H_x\tilde \nu_{\alpha} \tilde \nu_{\beta}} \tilde \nu_{\alpha}^* 
\tilde \nu_{\beta}
+ g_{H_x\tilde l_{\alpha} \tilde l_{\beta}} \tilde l_{\alpha}^* \tilde l_{\beta}
\right]\,\,,\nonumber\\
\mathcal{L}_{H_x \tilde \chi_i^- \tilde \chi_j^-}&=&
- g H_x\bar{\tilde{\chi}}_i^- 
\left[ W_{Lij}^{(x)}P_L+ W_{Rij}^{(x)}P_R \right] \tilde{\chi}_j^-\,\,,\nonumber\\
\mathcal{L}_{H_x \tilde \chi_a^0 \tilde \chi_b^0}&=&
- \frac{g}{2} H_x\bar{\tilde{ \chi}}_a^0 
\left[ D_{Lab}^{(x)}P_L+ D_{Rab}^{(x)}P_R \right] \tilde {\chi}_b^0\,\,,
\end{eqnarray}
where the coupling factors    
$S_{L,q}^{(x)}$, $S_{R,q}^{(x)}$, 
$g_{H_x\tilde \nu_{\alpha} \tilde \nu_{\beta}}$, $g_{H_x\tilde l_{\alpha} \tilde l_{\beta}}$,
$W_{Lij}^{(x)}$, $W_{Rij}^{(x)}$, $D_{Lab}^{(x)}$ and $D_{Rab}^{(x)}$ are collected in  Appendix B.

\subsection{LFVHD rates in the MSSM-seesaw}
\begin{figure}
\includegraphics[width=16.0cm,height=10.0cm]{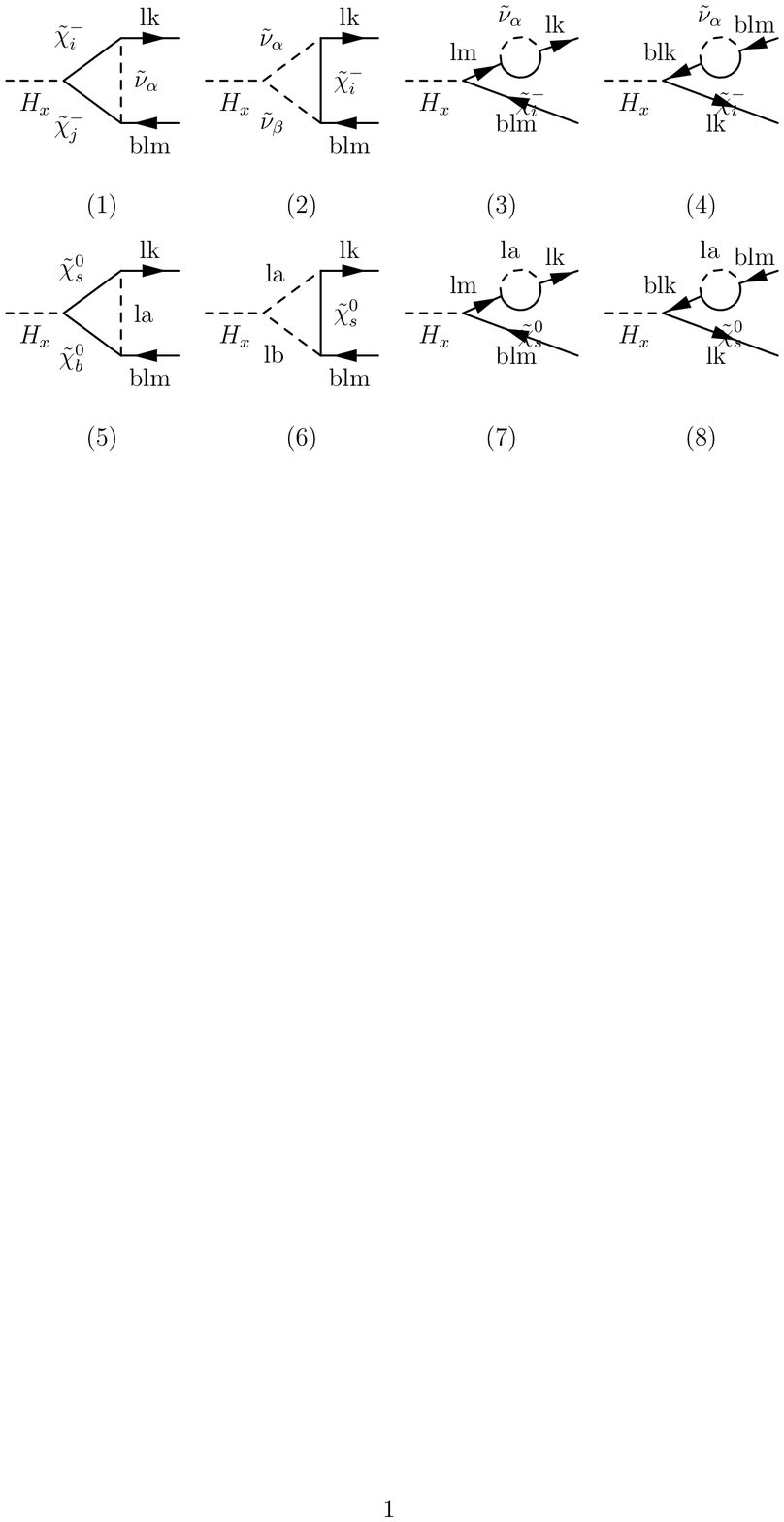}
\caption{One-loop diagrams for the LFVHD in the MSSM-seesaw model }
\label{fig:5}
\end{figure}

As we have said, the contributions to the LFVHD rates in the MSSM-seesaw come 
from various sectors.
 The contributions from the charged Higgs sector and from 
 the SM sector (i.e, $H^{\pm}$, $W^{\pm}$  and 
 $G^{\pm}$) are very small and will not be included here. The main contributions come from the 
 genuine SUSY sector, concretely, from the one-loop diagrams with charginos, neutralinos, 
 sleptons and sneutrinos shown in fig.~\ref{fig:5}.
 
%\vspace{1cm}
     
%%%%%%%%%%%%%%%%%%%%%%%%%%%%%%%%%%%%%%%%%%%%%%%%%%%%

%%%%%%%%%%%%%%%%%%%%%%%%%%%%%%%%%%%%%%%%%%%%%%%%%%%

The contributions of these one-loop diagrams to the form factors are given by,
\begin{equation}
F_{L,x} = \sum_{i=1}^{8} F_{L,x}^{(i)},\,\,  F_{R,x} = \sum_{i=1}^{8} F_{R,x}^{(i)},
\end{equation}
where the analytical results for $F_{L,x}^{(i)}$ and , $F_{R,x}^{(i)}$, $i=1..8$, are 
collected in Appendix B. 

\begin{figure}
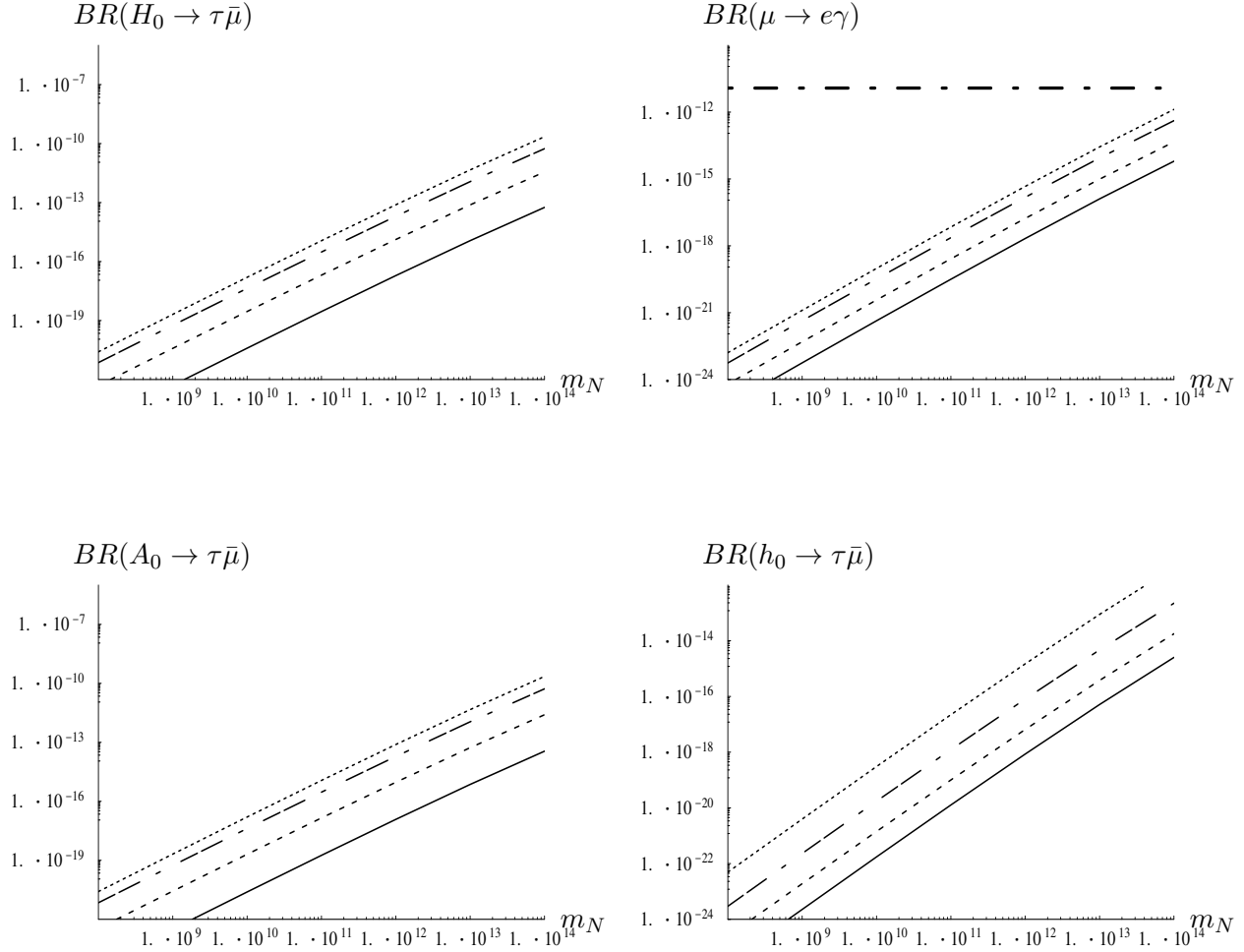

\hspace{-0.5cm}
\includegraphics[width=8.0cm,height=6.0cm]{fig6a.epsi}
\hspace{0.25cm}
\includegraphics[width=8.0cm,height=6.0cm]{fig6b.epsi}
\hspace{0.25cm}
\vspace{0.5cm}\\
\hspace{-0.5cm}
\includegraphics[width=8.0cm,height=6.0cm]{fig6c.epsi}
\hspace{0.25cm}
\includegraphics[width=8.0cm,height=6.0cm]{fig6d.epsi}
\caption{Dependence of $BR(H_x \to \tau \bar \mu)$ with $m_N$ (GeV) 
in scenario A with degenerate heavy neutrinos and real R,  for several values of 
$\tan\beta$. {\bf (6a)} Upper-left panel, $H_x=H_0$, {\bf (6c)} lower-left panel, $H_x=A_0$ and {\bf (6d)} lower-right panel, $H_x=h_0$. 
 {\bf (6b)} Upper-right panel: Dependence of $BR(\mu \to e \gamma)$ with $m_N$ 
 for several values of $\tan\beta$. In all plots,
the solid, dashed, dashed-dotted and  dotted lines are the preditions for  
$\tan \beta=3,10,30$ and $50$, respectively. The horizontal line in (6b) 
is the upper experimental bound on $BR(\mu \to e \gamma)$. 
The other input parameters are,
$M_0=400 $ GeV and $M_{1/2}=300 $ GeV.
\label{fig:6}}
\end{figure}

\begin{figure}
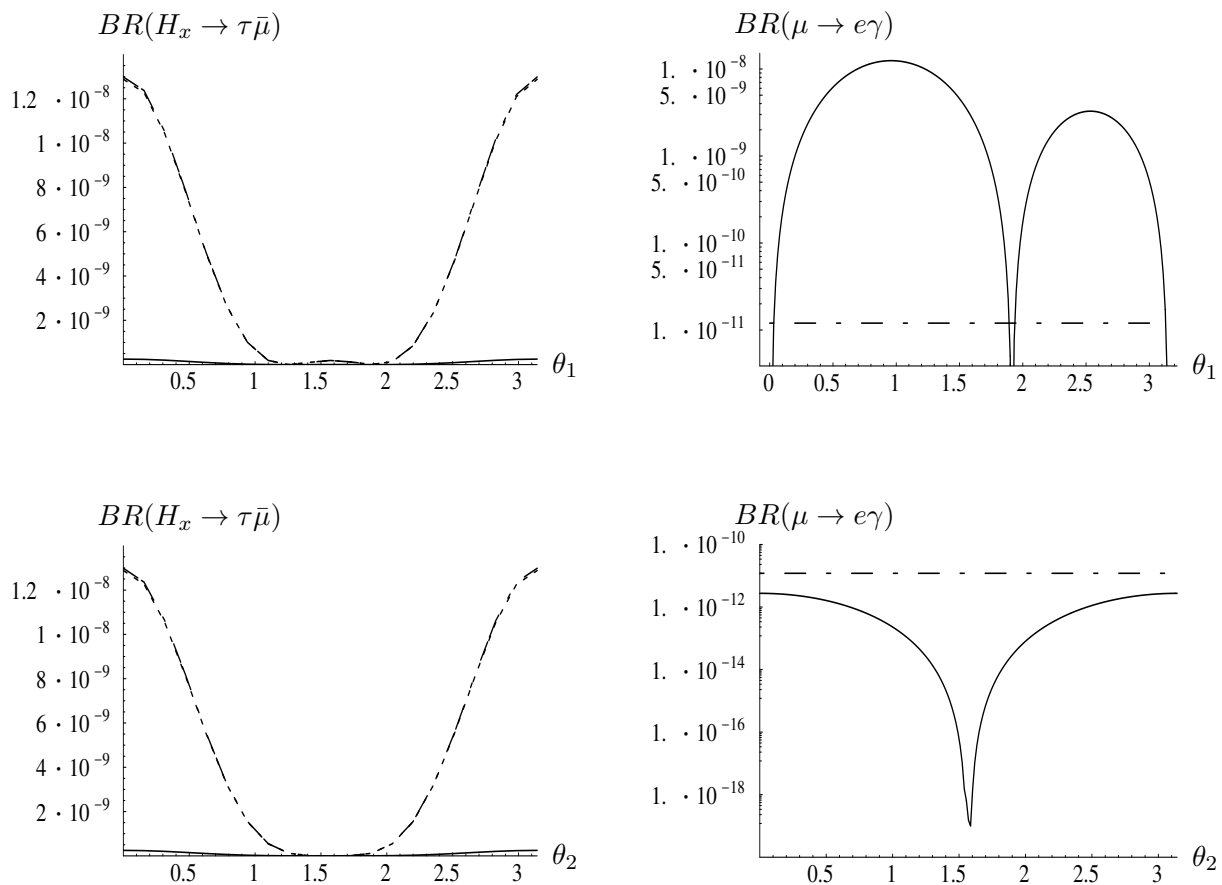

\hspace{-0.5cm}
\includegraphics[width=8.0cm,height=6.0cm]{fig7a.epsi}
\hspace{0.25cm}
\includegraphics[width=8.0cm,height=6.0cm]{fig7b.epsi}

\vspace{0.5cm}
\hspace{-0.5cm}
\includegraphics[width=8.0cm,height=6.0cm]{fig7c.epsi}
\hspace{0.25cm}
\includegraphics[width=8.0cm,height=6.0cm]{fig7d.epsi}
\caption{{\bf (7a)} Left panel: Dependence of $BR(H_x \to \tau \bar \mu)$ with  $\theta_1$. 
Solid, dashed and dashed-dotted lines (the two later undistinguishible here) correspond to $H_x=(h_0, H_0, A_0)$ respectively. 
{\bf (7b)} Right panel: Dependence of $BR(\mu \to e \gamma)$ with $\theta_1$. The horizontal dashed-dotted line is the upper experimental bound.
Both panels are in scenario B, case 1 for real $\theta_1 \neq 0$,
$(m_{N_1},m_{N_2},m_{N_3})=(10^8, 2 \times 10^8, 10^{14})$ GeV, $\theta_2=\theta_3=0$,
$\tan \beta = 50$, $M_0=400 $ GeV, and $M_{1/2}=300 $ GeV. {\bf (7c)}, 
lower left panel,
and {\bf (7d)}, lower right panel, 
are as in (7a) and (7b) respectively, but for $\theta_2 \neq 0$ and
$\theta_1=\theta_3=0$
\label{fig:7}}
\end{figure}

\begin{figure}
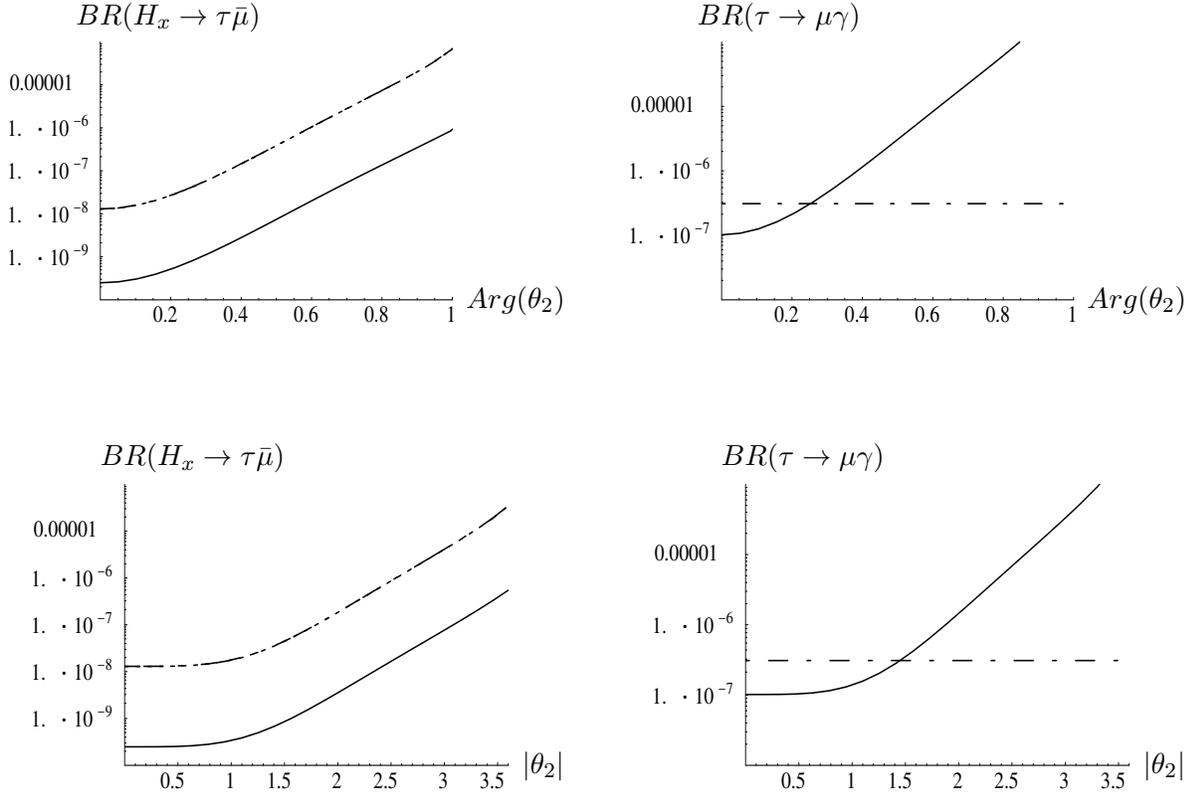

\includegraphics[width=7.5cm,height=5.5cm]{fig8a.epsi}
\hspace{0.5cm}
\includegraphics[width=7.5cm,height=5.5cm]{fig8b.epsi}
\hspace{0.5cm}

\vspace{0.5cm}
\includegraphics[width=7.5cm,height=5.5cm]{fig8c.epsi}
\hspace{0.5cm}
\includegraphics[width=7.5cm,height=5.5cm]{fig8d.epsi}
\caption{{\bf (8a)} Upper-left panel: Dependence of $BR(H_x \to \tau \bar \mu)$ with 
$Arg(\theta_2)$ for $|\theta_2|=\pi$.
{\bf (8b)} 
Upper-right panel: Dependence of $BR(\tau \to \mu \gamma)$ with $Arg(\theta_2)$ 
for $|\theta_2|=\pi$. {\bf (8c)} Lower-left panel: Dependence of $BR(H_x \to \tau \bar \mu)$ with 
$|\theta_2|$ for $Arg(\theta_2)=\pi/4$. {\bf (8d)} Lower-right panel: 
Dependence of $BR(\tau \to \mu \gamma)$ with$|\theta_2|$ for 
$Arg(\theta_2)=\pi/4$.
All figures are in scenario B, case 1, for complex $\theta_2$ and
$\theta_1=\theta_3=0$. The rest of parameters are fixed to:
$(m_{N_1},m_{N_2},m_{N_3})=(10^{8},2 \times 10^{8}, 10^{14})$ GeV,  
$\tan \beta = 50$, $M_0=400 $ GeV and  $M_{1/2}=300 $ GeV. Solid, dashed and dashed-dotted (the two later undistinguishible here) 
lines in the left panels correspond to $H_x=(h_0, H_0, A_0)$ 
respectively. The horizontal line in the right panels is the experimental upper
bound on $\tau \to \mu \gamma$.
\label{fig:8}}
\end{figure}

\begin{figure}
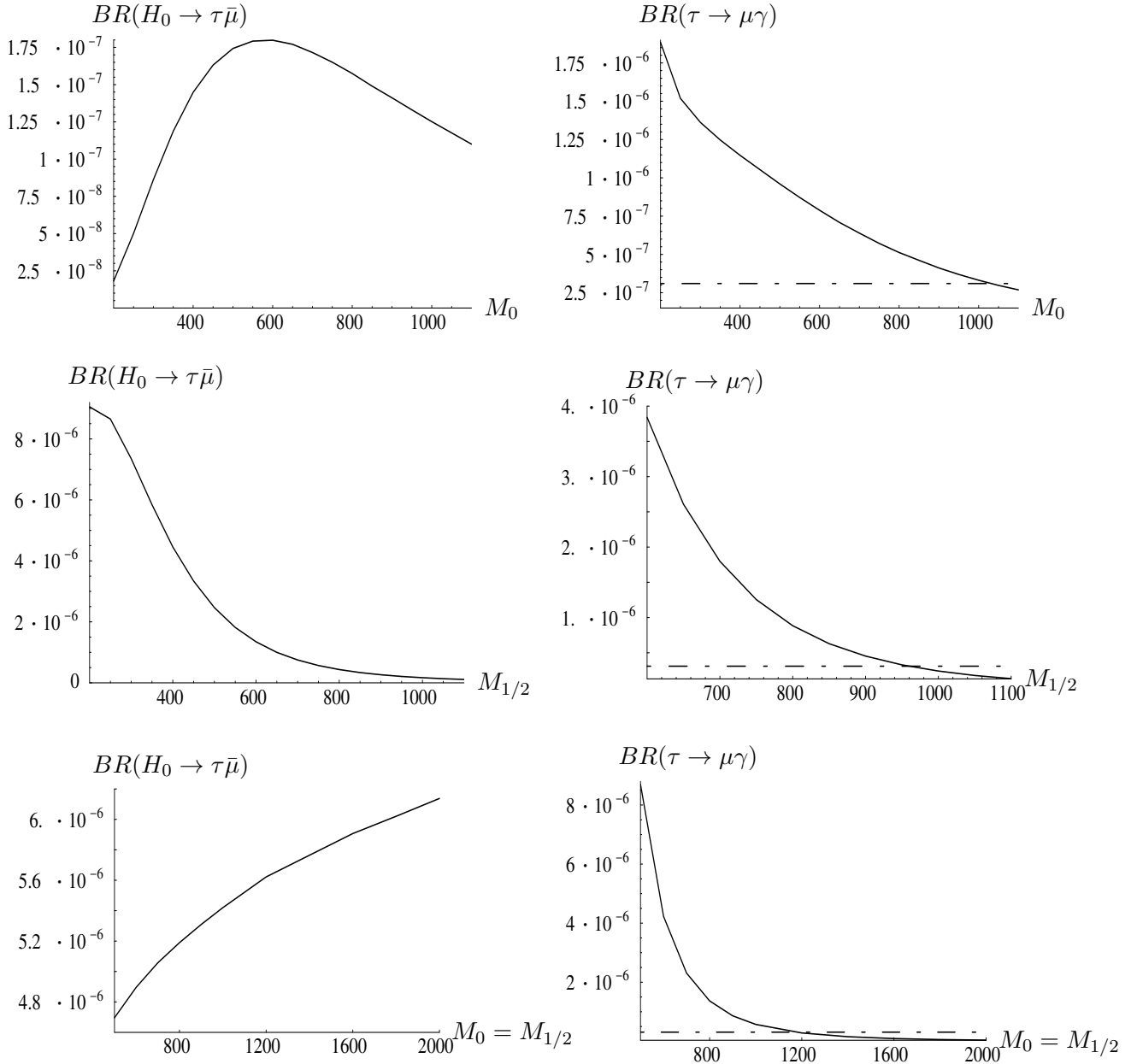

\hspace{-0.75cm}
\includegraphics[width=8.0cm,height=6.0cm]{fig9a.epsi}
\hspace{0.25cm}
\includegraphics[width=8.0cm,height=6.0cm]{fig9b.epsi}
\vspace{-0.3cm}
\\
\hspace{-0.5cm}
\includegraphics[width=8.0cm,height=6.0cm]{fig9c.epsi}
\hspace{0.25cm}
\includegraphics[width=8.0cm,height=6.0cm]{fig9d.epsi}
%\hspace{0.5cm}
\vspace{-0.3cm}
\\
\hspace{-0.5cm}
\includegraphics[width=8.0cm,height=6.0cm]{fig9e.epsi}
\hspace{0.25cm}
\includegraphics[width=8.0cm,height=6.0cm]{fig9f.epsi}
\caption{Dependence with $M_0$ (GeV) and $M_{1/2}$ (GeV) for scenario B 
with $(m_{N_1},m_{N_2},m_{N_3})=(10^{8},2 \times 10^{8}, 10^{14})$ GeV, $\theta_1=\theta_3=0$ and 
$\tan \beta = 50$.
{\bf (9a)} Upper left panel: Behaviour of $BR(H_0 \to \tau \bar \mu)$ with $M_0(GeV)$ 
for 
$M_{1/2}=300$ GeV and $\theta_2 =\pi e^{0.4i}$. 
{\bf (9b)} Upper right panel: Same as (9a) but for $BR(\tau \to \mu \gamma)$. 
{\bf (9c)} Medium left panel: Behaviour of $BR(H_0 \to \tau \bar \mu)$ with 
$M_{1/2}(GeV)$ for
$M_0=400$ GeV and $\theta_2 =\pi e^{0.8i}$. 
{\bf (9d)} Medium right panel: Same as (9c) but for $BR(\tau \to \mu \gamma)$. 
{\bf (9e)} Lower left panel: Behaviour of $BR(H_0 \to \tau \bar \mu)$ with 
$M_0=M_{1/2}(GeV)$ for $\theta_2 =\pi e^{0.8i}$. 
{\bf (9f)} Lower right panel: Same as (9e) but for $BR(\tau \to \mu \gamma)$.  
The horizontal line in the right panels is the upper experimental bound on $BR(\tau \to \mu \gamma)$. 
\label{fig:9}}
\end{figure}

\begin{figure}[h]
\includegraphics[width=7.5cm,height=5.5cm]{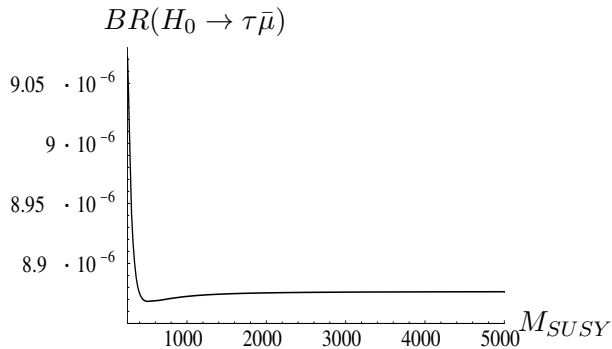}
\caption{Behaviour of $H_0 \to \tau \bar \mu$ in a generic MSSM scenario as a function of the 
common SUSY mass, $M_{SUSY} (GeV)\equiv m_{\tilde{L}, l}=m_{\tilde{E}, l}=M_0=\mu$. The gaugino soft masses are set to
$M_2=2/3 \mu$ and $M_1=\frac{5}{3}\tan^2\theta_WM_2$. Here we fix 
 $\lambda_{23}=-0.4$, $\lambda_{12}=\lambda_{13}=0$, $\tan\beta=50$ and 
 $m_{H_0}=340$ GeV. 
 \label{fig:10}}
\end{figure}

The partial widths for the $h^0,H^0,A^0 \to l_k \bar{l}_m$ decays are then finally computed by 
inserting these form factors correspondingly into eq.~\ref{decay}. 
We show in figs.~\ref{fig:6} to~\ref{fig:10}
the numerical results of the branching  ratios for the LFVHD in the MSSM. The   
total MSSM Higgs boson widths have been computed with the HDECAY 
programme~\cite{Djouadi:1997yw}. We have shown in the plots just the dominant 
channels, which are  $H_{x} \to \tau {\bar \mu}$, and some comments will be
added on the other channels. Similarly, for the comparison with 
the leptonic radiative decays, $l_j\to l_i \gamma$, we will show in the plots 
the most
relevant one, which is $\mu \to e \gamma$ or $\tau \to \mu \gamma$,
depending on the case. For the numerical estimates of 
the $l_j\to l_i \gamma$ branching ratios we use the exact analytical formulas of 
ref.~\cite{Hisano:1995cp}. These are expressed in the mass eigenstate basis
as well,  and contain all contributing one-loop diagrams. For the involved 
$\chi^-l{\tilde \nu}$ and $\chi^0 l{\tilde l}$ couplings we use again the expressions
of Appendix B.

The results of the branching ratios for the LFVHD, in the $\tau {\bar\mu}$ channel,
as a function of the Majorana
mass,  $m_N$, in scenario A with degenerate heavy neutrinos and real R, are illustrated in 
fig.~\ref{fig:6}, for several $\tan\beta$ values, $\tan\beta=3,10,30,50$. 
The explored range in $m_N$ is
from $10^8 $ GeV up to $10^{14} $ GeV which is favorable  
for baryogenesis. We also show in this figure, the
corresponding 
predicted rates for the most relevant lepton decay,
 which in this case is $\mu \to e \gamma$, and include its upper
 experimental bound. We have checked that the other channels are well within
 their experimental allowed range.   
From this figure we first see that 
the branching ratios for the light Higgs boson are smaller than the 
heavy Higgs ones in  about two orders of magnitude. 
The ratios of $H_0$ and $A_0$ are very similar in all the plots and, for this scenario,
they can reach values up to just $2.2 \times 10^{-10}$ in the region of high $\tan \beta$ and high $m_N$.
Besides, the rates for $\mu \to e \gamma$ decays are below the upper
experimental bound for all explored $\tan\beta$ and $m_N$ values. 
 From these plots  we also see clearly the  high sensitivity 
to $\tan\beta$ of the LFVHD rates for all Higgs
bosons which, at large $\tan \beta$, scale roughly as $(\tan\beta)^4$, in comparison
with the lepton decay rates which scale as $(\tan\beta)^2$. The dependence of both rates
on $m_N$ is that induced from the $\lambda_{ij}$ dependence, and corresponds approximately
to what is expected from the mass insertion approximation, where 
$BR(H_{x}\to l_j {\bar l_i})$, $BR (l_j \to l_i \gamma) 
\propto |\lambda_{ij}|^2 \propto |m_N\log(m_N)|^2 $. 

In what regards to the relative importance of the various SUSY sectors to the LFVHD
rates,     
we have found that these are dominated by the chargino contributions, that is 
from the loop diagrams (1),(2),(3) and (4) in fig.~\ref{fig:5}. For instance, for 
$m_N=10^{14}$ GeV , $M_0= 400 $ GeV, and $M_{1/2}=300 $ GeV, we have found the 
following ratios
between the chargino and neutralino contributions to the $H^0$ form factors: 
$|F_L^{\tilde\chi^-}/F_L^{\tilde\chi^0}|=6.1,\, 6,\,7.3,\,23.1$ for
$\tan\beta=3,\,10,\,30,\,50$ respectively,  where we have used a simplified notation, 
$F_L^{\tilde\chi^-}=F_{L,H_0}^{(1)}+F_{L,H_0}^{(2)}+F_{L,H_0}^{(3)}+F_{L,H_0}^{(4)}$,
$F_L^{\tilde\chi^0}=F_{L,H_0}^{(5)}+F_{L,H_0}^{(6)}+F_{L,H_0}^{(7)}+F_{L,H_0}^{(8)}$.
Similar $\tilde\chi^-/\tilde\chi^0$ ratios are found for the corresponding $F_R$ form
factors. The relative ratio found of $F_L/F_R \simeq 17$ is nicely explained by the 
$m_{\tau}/m_{\mu}$ ratio. For the lightest Higgs boson, we find 
$|F_L^{\tilde\chi^-}/F_L^{\tilde\chi^0}|=1.5,\, 1.4,\,1.7,\,4$ correspondingly. 

Concerning to the comparative size of the contributions from the 
various chargino loop diagrams we have found that, at large $\tan\beta$, the external 
leg corrections are clearly the dominant ones. Concretely, for 
$|(F_{L,H_0}^{(3)}+F_{L,H_0}^{(4)})/F_L^{\tilde\chi^-}|$, 
$|F_{L,H_0}^{(1)}/F_L^{\tilde\chi^-}|$ and $|F_{L,H_0}^{(2)}/F_L^{\tilde\chi^-}|$, we get the
respective percentages, 60.6\%, 39.3\% and 0.1\%, for $\tan\beta=10$ and 93.8\%, 6.2\%, 0\%
for $\tan\beta=50$. 

The branching ratios for the Higgs boson decays into $\tau\bar{e}$ and $\mu \bar e$ are much smaller
than the $\tau\bar{\mu}$ ones, as expected,  and we do not show plots for them. For instance, for 
$m_N=10^{14}$ GeV,
and $\tan\beta=50$ we find $BR(H^{(x)} \to \tau {\bar\mu})/BR(H^{(x)} \to \tau {\bar e})=3.9 \times 10^3$ and 
$BR(H^{(x)} \to \tau {\bar\mu})/BR(H^{(x)} \to \mu {\bar e})=1.3 \times 10^6$ for the three Higgs bosons.
 
All the previous results are for fixed $M_0=400$ GeV and $M_{1/2}=300$ GeV. The dependence with $M_0$ and $M_{1/2}$ will be discussed later on within the context of hierarchical neutrinos.

In summary, the LFVHD rates for degenerate heavy neutrinos are very small, at most 
  $2.2 \times 10^{-10}$, for the
  explored range of the seesaw parameters and $\tan \beta$.  
  Obviously, larger values of these LFVHD ratios 
 could be got for larger $\tan\beta$ values, but we have not considered them here.

We next present the results for hierarchical neutrinos, scenario B, and use the 
parametrization of case 1. The results
for real and complex $R$ and for the mass hierarchy  
$(m_{N_1},m_{N_2},m_{N_3})=(10^8, 2 \times 10^8, 10^{14})$ GeV, are shown
 in figs.~\ref{fig:7} and~\ref{fig:8} respectively. From these figures we first
 confirm that the LFVHD and lepton decay rates are larger in this case than in 
the degenerate heavy neutrinos one. However, we will get restrictions on 
the maximum allowed Higgs decay rates coming from the experimental lepton 
decay bounds.
For instance, the case of real $\theta_1$, that is illustrated in figures
~(\ref{fig:7}a) and~(\ref{fig:7}b) for $\tan \beta =50$, $M_0=400$ GeV and
$M_{1/2}=300$ GeV,
shows that compatibility with $\mu \to e \gamma$ data occurs
 only in the very narrow deeps at  around
 $\theta_1=0$, $1.9$ and $\pi$. The
 presence of these narrow regions where the $\mu \to e \gamma$ rates are
 drastically suppresed were already pointed out in ref.~\cite{Casas:2001sr} 
 and correspond clearly to the minima of $|\lambda_{12}|$ in 
 fig.~(\ref{fig:3}c). Notice that 
it is precisely at the points $\theta_1=0, \pi$ where the  
$BR (H_0, A_0 \to \tau \bar \mu)$ rates
reach their maximum values, although these are not large, just about $1.3 \times 10^{-8}$. Notice also, that 
these maxima correspond clearly to the maxima of $|\lambda_{23}|$ in 
fig.~(\ref{fig:3}c).
We have checked that for lower $\tan\beta$ values, the allowed regions in $\theta_1$ widen and are placed at the same points, but the corresponding maximum values of the LFVHD rates get considerably reduced. 
The alternative case of real $\theta_2 \neq 0$, with $\theta_1=\theta_3=0$ is illustrated in 
figs.~(\ref{fig:7}c) and ~(\ref{fig:7}d). We see that the behaviour of   
$BR (H_x \to \tau \bar \mu)$ with $\theta_2$ is very similar to that with  $\theta_1$ of 
fig.~(\ref{fig:7}a) and the maximum values of about $1.3 \times 10^{-8}$ are now placed at 
$\theta_2=0$, $\pi$.  $BR(\mu \to e \gamma)$  also reaches its maximum at
$\theta_2=0,\pi$, but it is still well below the experimental bound. 
In particular, for $\tan \beta =50$, $M_0=400$ GeV and
$M_{1/2}=300$ GeV this maximum value is $3 \times 10^{-12}$. Notice, that the behaviour with $\theta_2$ is 
explained once again in terms of the corresponding $\lambda_{ij}$ behaviour. 
Regarding the dependence with $\theta_3$, not shown in the plots, a different situation 
is found, where  $BR (H_x \to \tau \bar \mu)$ is approximately constant, and for the 
heavy Higgs bosons it is around $ 1.3 \times 10^{-8}$. 
$BR(\mu \to e \gamma)$,  $BR(\tau \to \mu \gamma)$ and $BR(\tau \to e \gamma)$ are also
approximately constant with $\theta_3$. In addition, we have checked that these three 
leptonic constant decay rates are 
within the experimental allowed range. 
In summary, for real R we find that the maximum allowed LFVHD rates are 
at or below $1.3 \times 10^{-8}$. 

The case of complex $R$ is certainly more promissing. The examples shown in 
figs.~(\ref{fig:8}a) and ~(\ref{fig:8}c) are for the most favourable case, among the ones
studied here, of complex $\theta_2 \neq 0$ with $\theta_1=\theta_3=0$. It shows that 
considerably larger $BR (H_x \to \tau \bar \mu)$ rates than in the real $R$ case 
are found. For the explored $\theta_2$ values in these plots, the Higgs rates 
grow with both $|\theta_2|$ and $Arg (\theta_2)$ and, for the selected values of the 
parameters 
in this figure, they reach values up to 
around $5 \times 10^{-5}$. 
We have checked that the predicted rates for $BR(\tau \to e
\gamma)$  are well below the experimental upper bound and that the $\mu \to e \gamma$ decay is, in
this case, less restrictive than the $\tau \to \mu \gamma$ decay.  
 Notice that the smallness of the $\mu \to e \gamma$ and $\tau \to e \gamma$ decay rates, 
 in the case under study of $\theta_2 \neq 0$, 
is not maintained if our hypothesis on $\theta_{13}=0$ is changed. For instance, for 
$\theta_{13}=5^o$, which is also allowed by neutrino data, 
we get $BR(\mu \to e \gamma)\sim 2.4 \times 10^{-8}$, for $\theta_2=\pi e^{i\frac{\pi}{10}}$, well above the
experimental upper bound. This is why we keep $\theta_{13}=0$ in all this work.
Therefore, in this case of complex $\theta_2 \neq 0$ with $\theta_{13}=0$, the relevant lepton decay is $\tau \to \mu \gamma$ which
is illustrated in figs.~(\ref{fig:8}b) and ~(\ref{fig:8}d) together with its experimental
bound. We see that the allowed region by $\tau \to \mu \gamma$ data of the $(|\theta_2|, Arg(\theta_2))$
parameter space implies a reduction in the Higgs rates, leading to a maximum 
allowed value of just $5 \times 10^{-8}$. 

These results are for fixed values of 
$(m_{N_1},m_{N_2},m_{N_3})=(10^8, 2 \times 10^8, 10^{14})$ GeV, $\tan\beta=50$, 
$M_0=400$ GeV and $M_{1/2}=300$ GeV. We have found that other choices of the 
soft SUSY breaking mass
parameters, $M_0$ and $M_{1/2}$, are more efficient in order to get 
larger maximum allowed
Higgs ratios. For instance, for $M_0=M_{1/2}=1200$ GeV, we find maximum allowed values 
of around $5.6 \times 10^{-6}$. The reason for this improvement is the different behaviour with these     
parameters of the LFVHD and the lepton decay rates, which will be studied in more detail 
next. 

In fig.~\ref{fig:9} we show the dependence of $BR(H^0\to\tau {\bar \mu})$ and 
$BR(\tau\to \mu \gamma)$ with $M_0$ and $M_{1/2}$ for hierarchical neutrinos with
$(m_{N_1},m_{N_2},m_{N_3})=(10^8, 2 \times 10^8, 10^{14})$ GeV and fixed values of
$\tan\beta=50$, $\theta_2 \neq 0$, and $\theta_1=\theta_3=0$. We see clearly 
in these plots the different behaviour of these two observables with the soft SUSY
breaking mass parameters. Figs.~(\ref{fig:9}a) and ~(\ref{fig:9}c) show a milder
depencence of $BR(H^0\to\tau {\bar \mu})$ on $M_0$ and $M_{1/2}$  than that 
of $BR(\tau \to \mu \gamma)$ in figs.~(\ref{fig:9}b) and~(\ref{fig:9}d) respectively.
This implies, that for large enough values of $M_0$ or $M_{1/2}$ or both the 
$BR(\tau \to \mu \gamma)$ rates get considerably suppresed, due to the decoupling of the heavy SUSY
particles in the loops, and enter into the allowed region by data, whereas the   
$BR(H^0\to\tau {\bar \mu})$ rates are not much reduced. In fact, we see in 
figs.~(\ref{fig:9}e) and ~(\ref{fig:9}f) that for the choice $M_0=M_{1/2}$ the tau 
decay ratio
crosses down the upper experimental bound at around $M_0=1200$ GeV whereas the Higgs decay ratio 
is still quite large $\sim 6 \times 10^{-6}$ in the high $M_0$ region, around $M_0 \simeq 2000$ GeV.  This behaviour with the soft SUSY breaking
parameters is a clear indication that the heavy SUSY particles in the loops do not
decouple in the LFVHD, in much the same way as it has been shown to happen 
in the case of Higgs decays into quarks with change of flavor~\cite{Curiel:2003uk}.
Notice that the non-decoupling of the SUSY particles in the LFVHD can also be reformulated as 
non-decoupling in the effective $H^{(x)} \tau \mu$ couplings and these in turn can induce large 
contributions to other LFV processes that are mediated by Higgs exchange as, for instance, 
$\tau \to \mu \mu \mu$~\cite{Babu:2002et}. However, we have checked that for the 
explored values in this work of $M_0$, $M_{1/2}$, $\tan \beta$, R and $m_{N_i}$ that lead to 
the anounced LFVHD ratios of about $6 \times 10^{-6}$, the corresponding 
$BR(\tau \to \mu \mu \mu)$ rates are below the present experimental upper bound of 
$2 \times 10^{-7}$~\cite{Yusa:2004gm}. 

Finally, in order to show more clearly the non-decoupling behaviour of the SUSY particles in 
 the contributing loops to the LFVHD 
 we consider, instead of mSUGRA, a simpler and more generic MSSM scenario, 
 with the $\lambda_{ij}$ being free parameters, 
 which we now fix
 to some particular values, concretely $\lambda_{23}=-0.4$, and $\lambda_{12}=\lambda_{13}=0$. For simplicity, 
 we also assume 
 a common SUSY mass at the electroweak scale, 
 $M_{SUSY}\equiv m_{\tilde{L}, l}=m_{\tilde{E}, l}=M_0=\mu$ and choose $M_2=\frac{2}{3} \mu$, 
 $M_1=\frac{5}{3}\tan^2\theta_WM_2$. This particular value of $\lambda_{23}$ corresponds roughly 
 to the predicted $\lambda_{23}$ in the MSSM-seesaw with the parameters set 
 in fig.~(\ref{fig:9}e).
 Finally, the $BR(H_0 \to \tau {\bar \mu})$ is shown in fig.~\ref{fig:10} as a function of this common 
 $M_{SUSY}$ scale, for $\tan\beta=50$ and $m_{H_0}=340$ GeV. 
 We see clearly that for large $M_{SUSY}$ the branching
 ratio approaches to a constant non-vanishing value, which for
 these input parameter values is
 of about $10^{-5}$, and therefore 
 the charginos, neutralinos, charged sleptons and sneutrinos do no decouple in 
 this observable. Another way of seing this explicitely, is by the analytical 
 computation of this observable in the large SUSY masses limit. We have
 performed this computation in the simplest case where all the SUSY masses are 
 equal and got the following asymptotic limits for the dominant form factor $F_L$,
 in the regime of small $\lambda_{23}$ and large $\tan\beta$:
\begin{eqnarray}
F_{L\chi^{\pm}}^{(H^0)}& = &-\frac{\alpha}{4\pi\sin^2\theta_W}
\frac{m_\tau}{12m_W}\lambda_{23}
(\tan\beta)^2 \\
F_{L\chi^{0}}^{(H^0)} &=& -\frac{\alpha}{4\pi\sin^2\theta_W}
\frac{m_\tau}{24m_W}(1-3\tan^2\theta_W)\lambda_{23}
(\tan\beta)^2
\end{eqnarray}
From these simple expressions we can estimate quite easily the LFVHD ratios.
For instance, for the parameters chosen in fig.~\ref{fig:10}, we get 
$BR(H^0 \to \tau {\bar \mu})\simeq 3 \times 10^{-6}$ in reasonable agreement with our numerical result in this figure and in fig.~(\ref{fig:9}e). Notice also that this
asymptotic result agrees with the result from the effective lagrangian approach 
in ref.~\cite{Brignole:2003iv}.
 Finally,  it is worth mentioning  that this non-decoupling behaviour is in contrast with the behaviour of  
 $BR(\mu \to e \gamma)$, which scales as $(M_W/M_{SUSY})^4$, 
 and  explains the comparatively large LFVHD 
 rates found here.

\section{\label{sec:5} Conclusions}

In this paper we have studied in full detail the lepton flavor violating Higgs boson decays that  
are produced if the neutrinos get their masses via the seesaw mechanism. We have considered
the two most popular seesaw models with three generations, the SM-seesaw and the MSSM-seesaw.
Within the SM-seesaw we have found extremely small branching ratios which are explained in terms 
of the decoupling behaviour of the heavy Majorana neutrinos and the smallness of the light
neutrino masses. In the MSSM-seesaw we find, in contrast, branching ratios that are many 
orders of magnitude larger. The larger ratios found are for $H_0 \to \tau {\bar \mu}$ and $A_0 \to \tau {\bar \mu}$ decays with similar rates. After
exploring the dependence of the $H_0 \to \tau {\bar \mu}$ decay rates 
with all the involved parameters 
of the MSSM-seesaw, and by requiring 
compatibility with data of the correlated predictions for $\mu \to e \gamma$, $\tau \to e \gamma$
and $\tau \to \mu \gamma$  decays, we find that  $BR(H_0 \to \tau {\bar \mu})$  as large as 
$~10^{-5}$, for hierarchical neutrinos and large $M_{SUSY}$ can be reached. These ratios are mostly sensitive to $\tan\beta$, the heaviest neutrino mass $m_{N_3}$ and the complex angle $\theta_2$, which have been 
taken in the range $3<\tan\beta<50$, $10^8 GeV < m_{N_3} < 10^{14} GeV $ and $(|\theta_2|, Arg(\theta_2)) \leq (3.5,1)$ respectively. The largest allowed ratios found in this work of about $10^{-5}$ are for 
$\tan\beta=50$, $m_{N_3}=10^{14}$ GeV, large $M_{SUSY}$ in the TeV range and for our choice of $\theta_2=\pi e^{0.8i}$, $\theta_1=\theta_3=0$, but a more refined analysis of the full parameter space could lead to even larger rates. In particular, it is obvious that larger $\tan \beta$ values will enhance considerably the rates and lead to Higgs ratios closer to the future 
experimental reach of $10^{-4}$ at LHC~\cite{Cotti:2001fm} and $e^+e^-$ and $\mu^+\mu^-$ colliders~\cite{Kanemura:2004cn}, but we have not tried this because it would require to perform a resummation
of the large $\tan\beta$ contributions that is beyond the scope of this work.

\begin{acknowledgments}
We are indebted to Alberto Casas and Alejandro Ibarra for valuable discussions and for helping us
in the estimate of the RGE-running effects on the neutrino masses and mixings. 
We wish to acknowledge Andrea Brignole and Anna Rossi
for warning us of an error in the previous version of this work. We also thank Ana Teixeira for 
her help with the usage of the  mSUSPECT  package. We appreciate the updated information on LFV data that was provided to us by Eric Torrence. D. Temes thanks F.Boudjema for useful
comments and for reading the manuscript. E. Arganda and A.M.Curiel acknowledge
the Spanish Ministery 
of Science and Education (MEC) for finantial support by their FPU grants, AP2003-3776 and AP2001-0678
respectively. This work was supported by the Spanish MEC under project FPA2003-04597. 
%\dots
\end{acknowledgments}

\appendix

\section{}
We present here the analytical results for the form factors in the SM-seesaw, in the
Feynman 't Hooft gauge.
\begin{eqnarray*}
&F_L^{(1)}& = -\frac{g^2}{4 m_W^3} \frac{1}{16 \pi^2} B_{l_k n_i} B_{l_m n_j}^* \left\{m_{l_k} m_{n_j} \left[(m_{n_i} + m_{n_j}) \mbox{Re}\left(C_{n_i n_j}\right) + i(m_{n_j} - m_{n_i}) \mbox{Im}\left(C_{n_i n_j}\right) \right] \tilde{C}_0 \right. \\
&&+ (C_{12} - C_{11}) \left[(m_{n_i} + m_{n_j}) \mbox{Re}\left(C_{n_i n_j}\right) \left( -m_{l_k}^3 m_{n_j} - m_{n_i} m_{l_k} m_{l_m}^2 + m_{n_i} m_{n_j}^2 m_{l_k} + m_{n_i}^2 m_{n_j} m_{l_k} \right) \right. \\
&&+ \left. \left. i(m_{n_j} - m_{n_i}) \mbox{Im}\left(C_{n_i n_j}\right) \left( -m_{l_k}^3 m_{n_j} + m_{n_i} m_{l_k} m_{l_m}^2 - m_{n_i} m_{n_j}^2 m_{l_k} + m_{n_i}^2 m_{n_j} m_{l_k} \right) \right] \right\}\\
&F_R^{(1)}& = \frac{g^2}{4 m_W^3} \frac{1}{16 \pi^2} B_{l_k n_i} B_{l_m n_j}^* \left\{ m_{n_i} m_{l_m} \left[(m_{n_i} + m_{n_j}) \mbox{Re}\left(C_{n_i n_j}\right) - i(m_{n_j} - m_{n_i}) \mbox{Im}\left(C_{n_i n_j}\right) \right] \tilde{C}_0 \right. \\
&&+ C_{12} \left[(m_{n_i} + m_{n_j}) \mbox{Re}\left(C_{n_i n_j}\right) \left( m_{l_m}^3 m_{n_i} - m_{n_i} m_{n_j}^2 m_{l_m} - m_{n_i}^2 m_{n_j} m_{l_m} + m_{n_j} m_{l_k}^2 m_{l_m} \right) \right. \\
&&+ \left. \left. i(m_{n_j} - m_{n_i}) \mbox{Im}\left(C_{n_i n_j}\right) \left( -m_{l_m}^3 m_{n_i} + m_{n_i} m_{n_j}^2 m_{l_m} - m_{n_i}^2 m_{n_j} m_{l_m} + m_{n_j} m_{l_k}^2 m_{l_m} \right) \right] \right\}
\end{eqnarray*}
where $C_{11, 12} = C_{11, 12}(m_{l_k}^2, m_H^2, m_W^2, m_{n_i}^2, m_{n_j}^2)$ and $\tilde{C}_0 = \tilde{C}_0(m_{l_k}^2, m_H^2, m_{W }^2, m_{n_i}^2, m_{n_j}^2)$.
\begin{eqnarray*}
&F_L^{(2)}& = \frac{g^2}{2 m_W} \frac{1}{16 \pi^2} B_{l_k n_i} B_{l_m n_j}^* m_{l_k} \left\{-m_{n_j} \left[(m_{n_i} + m_{n_j}) \mbox{Re}\left(C_{n_i n_j}\right) + i(m_{n_j} - m_{n_i}) \mbox{Im}\left(C_{n_i n_j}\right) \right] C_0 \right. \\
&&+ \left. (C_{12} - C_{11}) \left[(m_{n_i} + m_{n_j})^2 \, \mbox{Re}\left(C_{n_i n_j}\right) + i(m_{n_j} - m_{n_i})^2 \, \mbox{Im}\left(C_{n_i n_j}\right) \right] \right\}\\
&F_R^{(2)}& = -\frac{g^2}{2 m_W} \frac{1}{16 \pi^2} B_{l_k n_i} B_{l_m n_j}^* m_{l_m} \left\{m_{n_i} \left[(m_{n_i} + m_{n_j}) \mbox{Re}\left(C_{n_i n_j}\right) - i(m_{n_j} - m_{n_i}) \mbox{Im}\left(C_{n_i n_j}\right) \right] C_0 \right. \\
&&+ \left. C_{12} \left[(m_{n_i} + m_{n_j})^2 \, \mbox{Re}\left(C_{n_i n_j}\right) + i(m_{n_j} - m_{n_i})^2 \, \mbox{Im}\left(C_{n_i n_j}\right) \right] \right\}
\end{eqnarray*}
where $C_{0, 11, 12} = C_{0, 11, 12}(m_{l_k}^2, m_H^2, m_W^2, m_{n_i}^2, m_{n_j}^2)$.
\begin{eqnarray*}
&F_L^{(3)}& = \frac{g^2}{16 \pi^2} B_{l_k n_i} B_{l_m n_i}^* m_{l_k} m_W \left(C_{11} - C_{12}\right)\\
&F_R^{(3)}& =  \frac{g^2}{16 \pi^2} B_{l_k n_i} B_{l_m n_i}^* m_{l_m} m_W C_{12}
\end{eqnarray*}
where $C_{11, 12} = C_{11, 12}(m_{l_k}^2, m_H^2, m_{n_i}^2, m_W^2, m_W^2)$.
\begin{eqnarray*}
&F_L^{(4)}& = -\frac{g^2}{4 m_W} \frac{1}{16 \pi^2} B_{l_k n_i} B_{l_m n_i}^* m_{l_k} \left\{ m_{l_m}^2 (C_{12} - 2 C_{11}) + m_{n_i}^2 (C_{11} - C_{12}) - m_{n_i}^2 C_0 \right\} \\
&F_R^{(4)}& = -\frac{g^2}{4 m_W} \frac{1}{16 \pi^2} B_{l_k n_i} B_{l_m n_i}^* m_{l_m} \left\{ \tilde{C}_0 + 2 m_{l_m}^2 C_{11} + m_{n_i}^2 C_{12} + (m_{l_k}^2 - 2 m_H^2) (C_{11} - C_{12}) + 2 m_{n_i}^2 C_0 \right\}
\end{eqnarray*}
where $C_{0, 11, 12} = C_{0, 11, 12}(m_{l_k}^2, m_H^2, m_{n_i}^2, m_W^2, m_W^2)$ and $\tilde{C}_0 = \tilde{C}_0(m_{l_k}^2, m_H^2, m_{n_i}^2, m_W^2, m_W^2)$.
\begin{eqnarray*}
&F_L^{(5)}& = -\frac{g^2}{4 m_W} \frac{1}{16 \pi^2} B_{l_k n_i} B_{l_m n_i}^* m_{l_k} \left\{\tilde{C}_0 + 2 m_{n_i}^2 C_0 + (m_{n_i}^2 + 2 m_{l_k}^2) C_{11} + (m_{l_m}^2 - m_{n_i}^2 - 2 m_H^2) C_{12} \right\} \\
&F_R^{(5)}& = \frac{g^2}{4 m_W} \frac{1}{16 \pi^2} B_{l_k n_i} B_{l_m n_i}^* m_{l_m} \left\{m_{n_i}^2 C_0 + m_{l_k}^2 C_{11} + (m_{l_k}^2 - m_{n_i}^2) C_{12} \right\}
\end{eqnarray*}
where $C_{0, 11, 12} = C_{0, 11, 12}(m_{l_k}^2, m_H^2, m_{n_i}^2, m_W^2, m_W^2)$ and $\tilde{C}_0 = \tilde{C}_0(m_{l_k}^2, m_H^2, m_{n_i}^2, m_W^2, m_W^2)$.
\begin{eqnarray*}
&F_L^{(6)}& = \frac{g^2}{4 m_W^3} \frac{1}{16 \pi^2} B_{l_k n_i} B_{l_m n_i}^* m_{l_k} m_H^2 \left\{m_{n_i}^2 (C_0 + C_{11}) + (m_{l_m}^2 - m_{n_i}^2) C_{12} \right\} \\
&F_R^{(6)}& = \frac{g^2}{4 m_W^3} \frac{1}{16 \pi^2} B_{l_k n_i} B_{l_m n_i}^* m_{l_m} m_H^2 \left\{m_{n_i}^2 (C_0 + C_{12}) + m_{l_k}^2 (C_{11} - C_{12})  \right\}
\end{eqnarray*}
where $C_{0, 11, 12} = C_{0, 11, 12}(m_{l_k}^2, m_H^2, m_{n_i}^2, m_W^2, m_W^2)$.
\begin{eqnarray*}
&F_L^{(7)}& = \frac{g^2}{2 m_W} \frac{1}{16 \pi^2} B_{l_k n_i} B_{l_m n_i}^* \frac{m_{l_m}^2 m_{l_k}}{m_{l_k}^2 - m_{l_m}^2} B_1 \\
&F_R^{(7)}& = \frac{g^2}{2 m_W} \frac{1}{16 \pi^2} B_{l_k n_i} B_{l_m n_i}^* \frac{m_{l_k}^2 m_{l_m}}{m_{l_k}^2 - m_{l_m}^2} B_1 \\
&F_L^{(8)}& = \frac{g^2}{4 m_W^3} \frac{1}{16 \pi^2} B_{l_k n_i} B_{l_m n_i}^* \frac{m_{l_k}}{m_{l_k}^2 - m_{l_m}^2} \left\{ m_{l_m}^2 (m_{l_k}^2 + m_{n_i}^2) B_1 + 2 m_{n_i}^2 m_{l_m}^2 B_0 \right\} \\
&F_R^{(8)}& = \frac{g^2}{4 m_W^3} \frac{1}{16 \pi^2} B_{l_k n_i} B_{l_m n_i}^* \frac{m_{l_m}}{m_{l_k}^2 - m_{l_m}^2} \left\{ m_{l_k}^2 (m_{l_m}^2 + m_{n_i}^2) B_1 + m_{n_i}^2 (m_{l_k}^2 + m_{l_m}^2) B_0 \right\}
\end{eqnarray*}
where $B_{0, 1} = B_{0, 1}(m_{l_k}^2, m_{n_i}^2, m_W^2)$.
\begin{eqnarray*}
&F_L^{(9)}& = \frac{g^2}{2 m_W} \frac{1}{16 \pi^2} B_{l_k n_i} B_{l_m n_i}^* \frac{m_{l_m}^2 m_{l_k}}{m_{l_m}^2 - m_{l_k}^2} B_1 \\
&F_R^{(9)}& = \frac{g^2}{2 m_W} \frac{1}{16 \pi^2} B_{l_k n_i} B_{l_m n_i}^* \frac{m_{l_k}^2 m_{l_m}}{m_{l_m}^2 - m_{l_k}^2} B_1 \\
&F_L^{(10)}& = \frac{g^2}{4 m_W^3} \frac{1}{16 \pi^2} B_{l_k n_i} B_{l_m n_i}^* \frac{m_{l_k}}{m_{l_m}^2 - m_{l_k}^2} \left\{ m_{l_m}^2 (m_{l_k}^2 + m_{n_i}^2) B_1 + m_{n_i}^2 (m_{l_k}^2 + m_{l_m}^2) B_0 \right\} \\
&F_R^{(10)}& = \frac{g^2}{4 m_W^3} \frac{1}{16 \pi^2} B_{l_k n_i} B_{l_m n_i}^* \frac{m_{l_m}}{m_{l_m}^2 - m_{l_k}^2} \left\{ m_{l_k}^2 (m_{l_m}^2 + m_{n_i}^2) B_1 + 2 m_{n_i}^2 m_{l_k}^2 B_0 \right\}
\end{eqnarray*}
where $B_{0, 1} = B_{0, 1}(m_{l_m}^2, m_{n_i}^2, m_W^2)$.
\[
\tilde{C}_0(p_2^2, p_1^2, m_1^2, m_2^2, m_3^2) \equiv B_0(p_1^2, m_2^2, m_3^2) + m_1^2 C_0(p_2^2, p_1^2, m_1^2, m_2^2, m_3^2)
\]
In all the previous formulas, summation over all indices are understood. These run as
 $i,j=1,..6$ for neutrinos, and $k,m=1,..3$, for charged leptons. 
\section{}
\subsection{\label{app:subsec}Couplings in the MSSM-seesaw}
We present here the coupling factors entering in the MSSM-seesaw formulas.
\begin{eqnarray*}
A_{L\alpha j}^{(e,\mu,\tau)}&=& -\frac{m_{e,\mu,\tau}}{\sqrt{2}m_W
  cos\beta}U_{j2}^*
  R_{(1,2,3)\alpha}^{(\nu)} \nn \\
A_{R\alpha j}^{(e,\mu,\tau)}&=& V_{j1}
  R_{(1,2,3)\alpha}^{(\nu)}
\nn \\
B_{L\alpha a}^{(e,\mu,\tau)}&=& \sqrt{2}\left[\frac{m_{e,\mu,\tau}}{2m_W
  cos\beta}N_{a3}^*R_{(1,3,5)\alpha}^{(l)} +\left[\sin\theta_W
  N_{a1}^{'*}-\frac{\sin^2\theta_W}{cos\theta_W}N_{a2}^{'*}\right]R_{(2,4,6)\alpha}^{(l)}\right]\nn
  \\
B_{R\alpha a}^{(e,\mu,\tau)}&=& 
 \sqrt{2}\left[\left(-\sin\theta_WN_{a1}^{'}-\frac{1}{\cos\theta_W}(\frac{1}{2}-\sin^2\theta_W)N_{a2}^{'}\right)
 R_{(1,3,5)\alpha}^{(l)}+
\frac{m_{e,\mu,\tau}}{2m_W \cos\beta}N_{a3}R_{(2,4,6)\alpha}^{(l)}\right]\nn \\
W_{Lij}^{(x)}&=&\frac{1}{\sqrt{2}}\left(-\sigma_1^{(x)}U_{j2}^*V_{i1}^*+
\sigma_2^{(x)}U_{j1}^*V_{i2}^*\right)\,\,;\,\,
W_{Rij}^{(x)}=\frac{1}{\sqrt{2}}\left(-\sigma_1^{(x)}U_{i2}V_{j1}+\sigma_2^{(x)}U_{i1}V_{j2}\right)\nn
  \\
D_{Lab}^{(x)}&=&\frac{1}{2\cos\theta_W}\left[(\sin\theta_W N_{b1}^*-\cos\theta_W N_{b2}^*)(\sigma_1^{(x)}N_{a3}^*+\sigma_2^{(x)}N_{a4}^*)
\right.\nn \\
&+&(\sin\theta_W N_{a1}^*-\cos\theta_W
  N_{a2}^*)(\sigma_1^{(x)}N_{b3}^*+\sigma_2^{(x)}N_{b4}^*)\left.\right] \,\,;\,\,
 D_{Rab}^{(x)}=D_{Lab}^{(x)*}\nn \\
S_{L,l}^{(x)} &=& - \frac{m_{l}}{2 m_W \cos\beta}\:\sigma_1^{(x)*}\,\,;\,\, 
S_{R,l}^{(x)} =S_{L, l}^{(x)*} \nn \\
g_{H_x \tilde{l}_{\alpha} \tilde{l}_{\beta}} &=& 
-i g \left[ g_{LL, e}^{(x)} R_{1\alpha}^{*(l)} R_{1\beta}^{(l)} + 
g_{RR, e}^{(x)} R_{2\alpha}^{*(l)} R_{2\beta}^{(l)} + g_{LR, e}^{(x)} R_{1\alpha}^{*(l)} R_{2\beta}^{(l)} + 
g_{RL, e}^{(x)} R_{2\alpha}^{*(l)} R_{1\beta}^{(l)}\right. \\
&+& g_{LL, \mu}^{(x)} R_{3\alpha}^{*(l)} R_{3\beta}^{(l)} + g_{RR, \mu}^{(x)} R_{4\alpha}^{*(l)} R_{4\beta}^{(l)} + g_{LR, \mu}^{(x)} R_{3\alpha}^{*(l)} R_{4\beta}^{(l)} + g_{RL, \mu}^{(x)} R_{4\alpha}^{*(l)} R_{3\beta}^{(l)} \\
&+& \left. g_{LL, \tau}^{(x)} R_{5\alpha}^{*(l)} R_{5\beta}^{(l)} + 
g_{RR, \tau}^{(x)} R_{6\alpha}^{*(l)} R_{6\beta}^{(l)} + 
g_{LR, \tau}^{(x)} R_{5\alpha}^{*(l)} R_{6\beta}^{(l)} + 
g_{RL, \tau}^{(x)} R_{6\alpha}^{*(l)} R_{5\beta}^{(l)} \right] \nn \\
g_{H_x \tilde{\nu}_{\alpha} \tilde{\nu}_{\beta}} &= &-i g \left[ g_{LL, \nu}^{(x)} 
R_{1\alpha}^{*(\nu)} R_{1\beta}^{(\nu)} + g_{LL, \nu}^{(x)} R_{2\alpha}^{*(\nu)} R_{2\beta}^{(\nu)} + 
g_{LL, \nu}^{(x)} R_{3\alpha}^{*(\nu)} R_{3\beta}^{(\nu)} \right]\\
g_{LL, l}^{(x)} &=&  \frac{M_Z}{\cos{\theta_W}} \sigma_3^{(x)}
\left( \frac{1}{2}- \sin^2{\theta_W} \right) + \frac{m_{l}^2}{M_W \cos{\beta}} \sigma_4^{(x)}\\
g_{RR, l}^{(x)} &=&  \frac{M_Z}{\cos{\theta_W}} \sigma_3^{(x)}
\left(  \sin^2{\theta_W} \right) + \frac{m_{l}^2}{M_W \cos{\beta}}  \sigma_4^{(x)}\\
g_{LR, l}^{(x)} &=& 
\left(-\sigma_1^{(x)}A_l-\sigma_5^{(x)}\mu\right)
\frac{m_{l}}{2 M_W \cos{\beta}} \,\,;\,\,
g_{RL, l}^{(x)} = g_{LR, l}^{(x)*} \\
g_{LL, \nu}^{(x)} &=& -\frac{M_Z}{2\cos{\theta_W}} \sigma_3^{(x)}
\end{eqnarray*}

\[
\textrm{where }
\sigma_1^{(x)} = \begin{pmatrix}
  \sin\alpha& \\
  -\cos\alpha& \\
  i \sin\beta& 
\end{pmatrix},\;\;
\sigma_2^{(x)}=\begin{pmatrix}
 \cos\alpha\\
 \sin\alpha\\
 -i\cos\beta
\end{pmatrix},\;\;
\sigma_3^{(x)}=\begin{pmatrix}
 \sin(\alpha+\beta)\\
 -\cos(\alpha+\beta)\\
 0
\end{pmatrix}
\]
\[
\textrm{and }
\sigma_4^{(x)}=\begin{pmatrix}
 -\sin\alpha\\
 \cos\alpha\\
 0
\end{pmatrix},\;\;
\sigma_5^{(x)}=\begin{pmatrix}
 \cos\alpha\\
 \sin\alpha\\
 i\cos\beta
\end{pmatrix},\;\;
\textrm{ for } 
x=\begin{pmatrix}
 h^0\\
 H^0\\
 A^0
\end{pmatrix} 
\textrm{ respectively. } 
\]
The matrices that rotate to the mass eigenstate basis are: $U$ and $V$ for charginos, $N$
for neutralinos, $R^{(l)}$ for charged sleptons and $R^{(\nu)}$ for sneutrinos.  $U,V$ and $N$
are taken from ref.~\cite{Gunion:1984yn}, $N'_{a1}=N_{a1}\cos\theta_W+N_{a2}\sin\theta_W$,
$N'_{a2}=-N_{a1}\sin\theta_W+N_{a2}\cos\theta_W$, and $R^{(l)}$ and $R^{(\nu)}$ are computed 
here by the diagonalization procedure presented in section IV.  
The various indices in the previous formulas run as follows: $i,j=1,2$ for charginos, $a,b=1,..,4$
for neutralinos, $\alpha,\beta=1,..,6$ for charged sleptons, $\alpha, \beta=1,..,3$ for sneutrinos,
and $l=e,\mu,\tau$ for charged leptons. Summation over all indices is understood.
\subsection{\label{app:subsec2}Form factors in  the MSSM-seesaw}
We present here the analytical results for the form factors in the MSSM-seesaw. 
\begin{eqnarray*}
F_{L,x}^{(1)} &=& - \frac{g^2}{16 \pi^2}\left[\left(B_0 + 
m_{\tilde {\nu}_{\alpha}}^2 C_0+m_{l_m}^2 C_{12}+m_{l_k}^2 (C_{11} - C_{12})\right)\,
\kappa_{L 1}^{x,\,\tilde \chi^-} \right.\nn \\
&&+m_{l_k} m_{l_m} \left(C_{11}+C_0\right)\kappa_{L 2}^{x, \tilde \chi^-}+
m_{l_k} m_{\tilde \chi _j^-} \left(C_{11}-C_{12}+C_0\right)\kappa_{L 3}^{x, \tilde \chi^-}\,+ 
m_{l_m} m_{\tilde \chi _j^-} C_{12}\,\kappa_{L 4}^{x,\tilde \chi^-}\nn \\
&& +m_{l_k} m_{\tilde \chi _i^-} \left(C_{11}-C_{12}\right)\kappa_{L 5}^{x,\tilde \chi^-} + 
m_{l_m} m_{\tilde \chi _i^-} \left(C_{12}+C_0\right)\kappa_{L 6}^{x,\tilde \chi^-}+ 
m_{\tilde \chi _i^-} m_{\tilde \chi _j^-}C_0\,\kappa_{L 7}^{x,\tilde \chi^-} \left.\right] \nn \\
F_{L,x}^{(2)} &=&
- \frac{igg_{H_x\tilde {\nu}_\alpha \tilde {\nu}_\beta}}{16\pi^2}
\left[-m_{l_k}(C_{11}-C_{12})\,\iota_{L 1}^{x,\tilde \chi^-}-
m_{l_m} C_{12}\, \iota_{L 2}^{x,\tilde \chi^-}+
m_{\tilde \chi^-_i}C_0\,\iota_{L 3}^{x,\tilde \chi^-}\right]\nn \\
F_{L,x}^{(3)} &=&
- \frac{S_{L,l_{m}}^{(x)}}{m_{l_k}^2-m_{l_m}^2}\left[m_{l_k}^2 \Sigma_R^{\tilde
    \chi^-} (m_{l_k}^2)+m_{l_k}^2 \Sigma_{Rs}^{\tilde \chi^-}(m_{l_k}^2)\right.
+ m_{l_m}\left(m_{l_k} \Sigma_L^{\tilde \chi^-} (m_{l_k}^2)+
m_{l_k}\Sigma_{Ls}^{\tilde \chi^-} (m_{l_k}^2)\right)\left.\right]\nn \\
F_{L,x}^{(4)} &=&
- \frac{S_{L,l_{k}}^{(x)}}{m_{l_m}^2-m_{l_k}^2}
\left[m_{l_m}^2 \Sigma_L^{\tilde \chi^-} (m_{l_m}^2)+
m_{l_m} m_{l_k} \Sigma_{Rs}^{\tilde \chi^-}(m_{l_m}^2)\right.
+ \left. m_{l_k}\left(m_{l_m} \Sigma_R^{\tilde \chi^-} (m_{l_m}^2)
+m_{l_k} \Sigma_{Ls}^{\tilde \chi^-} (m_{l_m}^2)\right)\right]\nn \\
F_{L,x}^{(5)} &=& - \frac{g^2}{16 \pi^2}\left[\left(B_0 + 
m_{\tilde l_{\alpha}}^2 C_0+m_{l_m}^2 C_{12}+m_{l_k}^2 (C_{11} - C_{12})\right)\,
\kappa_{L 1}^{x,\,\tilde \chi^0} \right.\nn \\
&&+m_{l_k} m_{l_m} \left(C_{11}+C_0\right)\kappa_{L 2}^{x, \tilde \chi^0}+
m_{l_k} m_{\tilde \chi _b^0} \left(C_{11}-C_{12}+C_0\right)\kappa_{L 3}^{x, \tilde \chi^0}\,+ 
m_{l_m} m_{\tilde \chi _b^0} C_{12}\,\kappa_{L 4}^{x,\tilde \chi^0}\nn \\
&& +m_{l_k} m_{\tilde \chi _a^0} \left(C_{11}-C_{12}\right)\kappa_{L 5}^{x,\tilde \chi^0} + 
m_{l_m} m_{\tilde \chi _a^0} \left(C_{12}+C_0\right)\kappa_{L 6}^{x,\tilde \chi^0}+ 
m_{\tilde \chi _a^0} m_{\tilde \chi _b^0}C_0\,\kappa_{L 7}^{x,\tilde \chi^0} \left.\right] \nn \\
F_{L,x}^{(6)} &=&
- \frac{igg_{H_x\tilde l_\alpha \tilde l_\beta}}{16\pi^2}
\left[-m_{l_k}(C_{11}-C_{12})\,\iota_{L 1}^{x,\tilde \chi^0}-
m_{l_m} C_{12}\, \iota_{L 2}^{x,\tilde \chi^0}+
m_{\tilde \chi_a^0}C_0\,\iota_{L 3}^{x,\tilde \chi^0}\right]\nn \\
F_{L,x}^{(7)} &=&
- \frac{S_{L,l_m}^{(x)}}{m_{l_k}^2-m_{l_m}^2}\left[m_{l_k}^2 \Sigma_R^{\tilde
    \chi^0} (m_{l_k}^2)+m_{l_k}^2 \Sigma_{Rs}^{\tilde \chi^0}(m_{l_k}^2)\right.
+ m_{l_m}\left(m_{l_k} \Sigma_L^{\tilde \chi^0} (m_{l_k}^2)+
m_{l_k}\Sigma_{Ls}^{\tilde \chi^0} (m_{l_k}^2)\right)\left.\right]\nn \\
F_{L,x}^{(8)} &=&
- \frac{S_{L,l_k}^{(x)}}{m_{l_m}^2-m_{l_k}^2}
\left[m_{l_m}^2 \Sigma_L^{\tilde \chi^0} (m_{l_m}^2)+
m_{l_m} m_{l_k} \Sigma_{Rs}^{\tilde \chi^0}(m_{l_m}^2)\right.
+ \left. m_{l_k}\left(m_{l_m} \Sigma_R^{\tilde \chi^0} (m_{l_m}^2)
+m_{l_k} \Sigma_{Ls}^{\tilde \chi^0} (m_{l_m}^2)\right)\right]\nn 
\label{formfactorLbs}
\end{eqnarray*}
where,
 \[C_{0,11,12}=
 \left\{   \begin{array}{l}
%\vspace{2em}
C_{0,11,12}
(m_{l_k}^2,m_{H_x}^2, m_{\tilde {\nu}_ {\alpha}}^2,m_{\tilde \chi^-_i}^2,m_{\tilde \chi^-_j}^2)
\textrm{ in } F_{L,x}^{(1)}\\
C_{0,11,12}
(m_{l_k}^2,m_{H_x}^2, m_{\tilde \chi^-_i}^2,m_{\tilde {\nu}_ {\alpha}}^2,m_{\tilde {\nu}_ {\beta}}^2)
\textrm{ in } F_{L,x}^{(2)}
  \\C_{0,11,12}
(m_{l_k}^2,m_{H_x}^2, m_{\tilde {l}_ {\alpha}}^2,m_{\tilde \chi^0_a}^2,m_{\tilde \chi^0_b}^2)
\textrm{ in } F_{L,x}^{(5)}
  \\C_{0,11,12}
(m_{l_k}^2,m_{H_x}^2, m_{\tilde \chi^0_a}^2,m_{\tilde {l}_ {\alpha}}^2,m_{\tilde {l}_ {\beta}}^2) 
\textrm{ in } F_{L,x}^{(6)}
\end{array}
\right. 
\]

and,

\begin{eqnarray} 
\Sigma^{\tilde \chi} (k) &=& k\, \Sigma_L^{\tilde \chi} (k^2) P_L + 
 k\,
\Sigma_R^{\tilde \chi} (k^2) P_R + m \left[ \Sigma_{Ls}^{\tilde \chi}(k^2) P_L + 
\Sigma_{Rs}^{\tilde \chi} (k^2) P_R \right]\, . 
\end{eqnarray}
The coupling factors and self-energies appearing in the neutralino contributions to the form
factors are given by, 
\begin{eqnarray*}
     \kappa_{L 1}^{x,\,\tilde \chi^0} = B_{L \alpha a}^{(l_k)}D_{R
     ab}^{(x)}B_{R \alpha b}^{(l_m)*} && \iota_{L 1}^{x, \tilde\chi^0} = 
     B_{R \alpha a}^{(l_k)}B_{R \beta a}^{(l_m)*}\\
     \kappa_{L 2}^{x,\,\tilde \chi^0} = B_{R \alpha a}^{(l_k)}D_{L
     ab}^{(x)}B_{L \alpha b}^{(l_m)*} && \iota_{L 2}^{x, \tilde\chi^0} = 
     B_{L \alpha a}^{(l_k)}B_{L \beta a}^{(l_m)*}\\
     \kappa_{L 3}^{x,\,\tilde \chi^0} = B_{R \alpha a}^{(l_k)}D_{L
     ab}^{(x)}B_{R \alpha b}^{(l_m)*} && \iota_{L 3}^{x, \tilde\chi^0} = 
     B_{L \alpha a}^{(l_k)}B_{R \beta a}^{(l_m)*}\\
     \kappa_{L 4}^{x,\,\tilde \chi^0} = B_{L \alpha a}^{(l_k)}D_{R
     ab}^{(x)}B_{L \alpha b}^{(l_m)*} && \\
     \kappa_{L 5}^{x,\,\tilde \chi^0} = B_{R \alpha a}^{(l_k)}D_{R ab}^{(x)}B_{R \alpha b}^{(l_m)*} &&\\
     \kappa_{L 6}^{x,\,\tilde \chi^0} = B_{L \alpha a}^{(l_k)}D_{L ab}^{(x)}B_{L \alpha b}^{(l_m)*} &&\\
     \kappa_{L 7}^{x,\,\tilde \chi^0} = B_{L \alpha a}^{(l_k)}D_{L
     ab}^{(x)}B_{R \alpha b}^{(l_m)*} &&
\end{eqnarray*}\vspace*{-0.7cm}
\begin{eqnarray}
\Sigma_L^{\tilde\chi^0} (k^2) &=&
-\frac{g^2}{16\pi^2}B_1(k^2,m_{\tilde\chi_a^0}^2, m_{\tilde l_\alpha}^2)
       B_{R\,\alpha a}^{(l_k)}B_{R\,\alpha a}^{(l_m)*} \nn \\
     m_{l_k} \Sigma_{Ls}^{\tilde \chi^0} (k^2) &=&  \frac{g^2m_{\tilde\chi_a^0}}{16\pi^2}B_0(k^2,
     m_{\tilde\chi_a^0}^2, m_{\tilde l_\alpha}^2)
       B_{L\,\alpha a}^{(l_k)}B_{R\,\alpha a}^{(l_m)*} 
\end{eqnarray}

The coupling factors and self-energies appearing in the chargino contributions to the form factors,  
 $\kappa^{x, \, \tilde\chi^-}$,
$\iota^{x, \, \tilde\chi^-}$, and $\Sigma^{\tilde\chi^-}$  can be
obtained from the previous expressions for
$\kappa^{x, \, \tilde\chi^0}$,  $\iota^{x, \, \tilde\chi^0}$ and $\Sigma^{ \tilde\chi^0}$ by making 
the replacements 
$m_{\tilde\chi_a^0}\rightarrow m_{\tilde\chi_i^-}$, $m_{\tilde l_\alpha}\rightarrow 
m_{\tilde \nu_\alpha}$, 
$B^{(l)}\rightarrow A^{(l)}$,  
$D^{(x)}\rightarrow W^{(x)}$, $a \rightarrow i$, and $b \rightarrow j$.
 
 The form factors $F_{R,x}^{(i)},i=1,..8$ can be got from $F_{L,x}^{(i)},i=1,..8$ by exchanging
 the indices $L\leftrightarrow R$ everywhere.
%\section{}
%
%\subsection{\label{app:subsec}A subsection in an appendix}
%
%
%You can use a subsection or subsubsection in an appendix. Note the
%numbering: we are now in Appendix \ref{app:subsec}.
%
%Note the equation numbers in this appendix, produced with the
%subequations environment:
%\begin{subequations}
%\begin{eqnarray}
%E&=&mc, \label{appa}
%\\
%E&=&mc^2, \label{appb}
%\\
%E&\agt& mc^3. \label{appc}
%\end{eqnarray}
%\end{subequations}
%
%They turn out to be Eqs.~(\ref{appa}), (\ref{appb}), and (\ref{appc}).
\newpage %Just because of unusual number of tables stacked at end

%%%%%%%%%%%%%%%%%%%%%%%%%%%%%%%%%%FIGURAS%%%%%%%%%%%%%%%%%%%%%%%%%%%%%%%%%%%%%%%%%%%%%%%%%%%%%55

\end{document}